\documentclass[review]{elsarticle}
\usepackage{hyperref}
\usepackage{mathtools}
\usepackage{float}
\usepackage{verbatim} 
\usepackage{apalike}
\usepackage[normalem]{ulem}

\usepackage{subcaption}
\usepackage{amsmath,amssymb,amsfonts}

\usepackage{multirow}

\newcommand{\nc}{\textcolor{red}}

\usepackage{booktabs}

\usepackage{graphicx}
 \usepackage{xcolor}
\usepackage[ruled,vlined]{algorithm2e}
\restylefloat{figure}

\journal{Expert Systems with Applications}

\biboptions{authoryear}

\begin{document}
\begin{frontmatter}

 \begin{titlepage}
 \begin{center}
 \vspace*{1cm}

 \textbf{ \large A Semi-Automated Hybrid Schema Matching Framework for Vegetation Data Integration}

 \vspace{1.5cm}

 Md Asif-Ur-Rahman$^a$ (rahman.m.asifur@gmail.com), Bayzid Ashik Hossain$^a$ (bayzid.hossain@det.nsw.edu.au), Michael Bewong$^b$ (mbewong@csu.edu.au), Md Zahidul Islam$^a$ (zislam@csu.edu.au), Yanchang Zhao$^c$ (yanchang.zhao@data61.csiro.au), Jeremy Groves$^d$ (jeremy.groves@cer.gov.au), Rory Judith$^d$ (rory.judith@environment.gov.au)  \\

 \hspace{10pt}

 \begin{flushleft}
 \small  
 $^a$ School of Computing, Mathematics and Engineering, Charles Sturt University, Panorama Avenue, Bathurst, NSW 2795, Australia. \\
 $^b$ School of Computing, Mathematics and Engineering, Charles Sturt University, Boorooma street, Locked Bag 588, Wagga Wagga, NSW 2678, Australia.\\
 $^c$ CSIRO Data61, GPO Box 1700, 
Canberra, ACT 2601, Australia.\\
$^d$ Department of Climate Change, Energy, the Environment and Water, GPO Box 3090, Canberra ACT 2601, Australia.


 \vspace{1cm}
 \textbf{Corresponding Author:} \\
 Michael Bewong \\
School of Computing, Mathematics and Engineering, Charles Sturt University, Boorooma street, Locked Bag 588, Wagga Wagga, NSW 2678, Australia. \\
 Tel: +61 02 6933 2591 \\
 Email: mbewong@csu.edu.au

 \end{flushleft}        
 \end{center}
 \end{titlepage}
\begin{abstract}
Integrating disparate and distributed vegetation data is critical for consistent and informed national policy development and management. Australia’s National Vegetation Information System (NVIS) under the Department of Climate Change, Energy, the Environment and Water (DCCEEW) is the only nationally consistent vegetation database and hierarchical typology of vegetation types in different locations. 
Currently, this database employs manual approaches for integrating disparate state and territory datasets which is labour intensive and can be prone to human errors. To cope with the ever-increasing need for up to date vegetation  data  derived from  heterogeneous data sources, a Semi-Automated Hybrid Matcher (SAHM) is proposed in this paper. SAHM utilizes both schema level and instance level matching following a two-tier matching framework. A key novel technique in SAHM called Multivariate Statistical Matching is proposed for automated schema scoring which takes advantage of domain knowledge and correlations between attributes to enhance the matching. To verify the effectiveness of the proposed framework, the performance of the individual as well as combined components of SAHM have been evaluated. The empirical evaluation shows the effectiveness of the proposed framework which outperforms existing state of the art methods like Cupid, Coma, Similarity Flooding, Jaccard Leven Matcher, Distribution Based Matcher, and EmbDI. In particular, SAHM achieves between 88\% and 100\% accuracy with significantly better F1 scores in comparison with state-of-the-art techniques. SAHM is also shown to be several orders of magnitude more efficient than existing techniques.
\end{abstract}


\end{frontmatter}

\section{Introduction}
\label{introduction}
The National Vegetation Information System (NVIS)\citep{nvismanual} is an ongoing collaborative initiative between the Australian federal, and state and territory governments. It is the only nationally comprehensive means of describing and representing vegetation information based on the relationships between structural and floristics\footnote{ The scientific study of plant species present in an area.} data. The NVIS is continuously updated for improved vegetation planning and management within Australia such as for events like bushfire. It also underpins a multitude of national and international legislation, policies and reporting. However, one of the main challenges for the NVIS is integrating heterogeneous datasets into the federal database. 

There is a strong impetus for data driven policy within the Commonwealth government \citep{impnvis1,impnvis2,impnvis3}. To provide these insights there is a need for the Commonwealth to rapidly integrate disparate datasets within the NVIS. With many new data sources, the Department’s need for data is rapidly increasing. Regrettably, schema matching remains essentially a manual, labour-intensive process that often requires a high level of domain knowledge. The effort needed is generally linear in the number of schemas to be matched with increasing complexity \citep{impofsm}. This limits the data flow into the department as the amount of resources needed to integrate new data sources is often very large. Automated schema matching would significantly ease this bottleneck within the department’s data collation and increase the department’s knowledge base.

Schema matching research has been ongoing for more than thirty years. Currently, researchers are turning towards the automation of schema matching for identification of correspondences among database attributes  \citep{i1,learning_to_rerank}. Existing techniques often use simple statistics like mean and standard deviation \citep{comapp1,smwithdd,statisticalsm}, and string based comparison such as prefix/suffix tests and edit distances~\citep{automatic_discovery_of_attribute,stringmatcher1,stringmatcher2,pc}. Nevertheless, the challenges with these approaches are that heterogeneous schemas which represent the same concepts may have different structural and naming formats. At the same time, different schemas may have identical structural and naming formats but representing very different information content. In addition, some attribute types (e.g., numerical) may  need to be mapped into different attribute types (e.g., categorical) during data integration. Unfortunately, existing schema or instance based matching techniques do not fully address these challenges.

In this paper, a novel hybrid schema matching framework is proposed for the automation of NVIS database integration. This work is funded by the Department of Climate Change, Energy, the Environment and Water (DCCEEW)\footnote{formerly Department of Agriculture, Water and the Environment (DAWE)}, and Data61 of CSIRO. To effectively automate the overall process, we propose SAHM, a two-tiered solution using both schema level and instance level matching that addresses the above challenges. The first tier, called \emph{schema level matcher} relies on the schema level information and comprises of two matchers, namely Domain Knowledge Matching and Linguistic Matching. The second tier, called \emph{instance level matcher} relies on the intance level information and comprises of two matchers, namely Univariate Statistical Matching and Multivariate Statistical Matching. In the proposed framework, domain knowledge matching incorporates domain experts’ knowledge in terms of a set of predefined rules; linguistic matching uses similarity based metrics on the attributes; univariate statistical matching uses the distribution of values within attributes; and multivariate statistical matching leverages the correlations among multiple attributes to ensure more accurate schema matching. 

While each of the various matching techniques contributes significantly to the effectiveness of SAHM, as demonstrated in the empirical section, the crux of SAHM is the novel multivariate statistical matching. This technique pivots around any known matching pairs in the source and destination datasets. That is, when a user selects an attribute from the source dataset denoted $a^s$ (e.g., state dataset) and then selects another attribute from the destination dataset denoted $a^d$ (e.g., NVIS database) known to be a match, the proposed novel technique calculates the correlation between the selected attribute and the rest of the attributes of the source dataset (resp. destination dataset). The technique then tries to find potential source-destination matching pairs based on the correlation scores of the remaining attributes wrt the selected source attribute (resp. destination attribute). We hypothesise that, if any source and destination attribute are a match, then they will bear a similar correlation to $a^s$ and $a^d$ respectively. Thus, exploiting this correlation, multivariate statistical matching scheme is capable of effectively identifying source and destination attribute matches.

The rest of this paper is organised as follows. Section 2 gives the background of this schema matching challenge and presents the related literature in schema matching. Section 3 presents the proposed solution by skillfully incoporating some state-of-the-art schema matching techniques such as Levenshtein similarity~\citep{levenshtein}, Monge-Elkan similarity~\citep{monge}, TF-IDF~\citep{tfidf} and univariate statistical matching with our novel multivariate statistical matching technique. Section 4 presents the effectiveness of the proposed framework by analyzing the performance of the individual matchers as well as the overall framework of SAHM. Finally Section 5 presents a conclusion.

\section{Background and Related Work}
In this section, we present the National Vegetation Information System (NVIS) schema matching environment. Next, we discuss the related literature in schema mapping, and analyse the challenges in applying existing techniques to the unique NVIS schema mapping environment. Table~\ref{tab:notations} summarises the frequently used notations in this paper.

\begin{table}[h]
\small
    \caption{Notations}
    \begin{tabular}{|c|c|}
    \hline
    \textbf{Symbol}  & \textbf{Meaning} \\
    \hline
    $R^{s_i}$, $a^{s_i}$,$n_{s_i}$  & source data, source attribute (state $i$), number of attributes in $R^{s_i}$ \\
   \hline
   $R^{d}$, $a^d$, $n_{d}$  & dest. data, dest. attribute (NVIS), number of attributes in $R^{d}$\\
    \hline
   $M^{sl}$, $score^{sl}$ & schema level matcher, schema level matching scores\\
   \hline
   $M^{il}$, $score^{il}$ & instance level matcher, instance level matching scores\\
   \hline
   $pc$ & pearson correlation score\\
    \hline
    $Sim_*$ & similarity function *\\
    \hline
    \end{tabular}
    \label{tab:notations}
\end{table}

\normalfont
\subsection{NVIS Schema Mapping Environment}
\begin{figure*}[t]
\centering
\includegraphics[width=4.9in,height=3in]{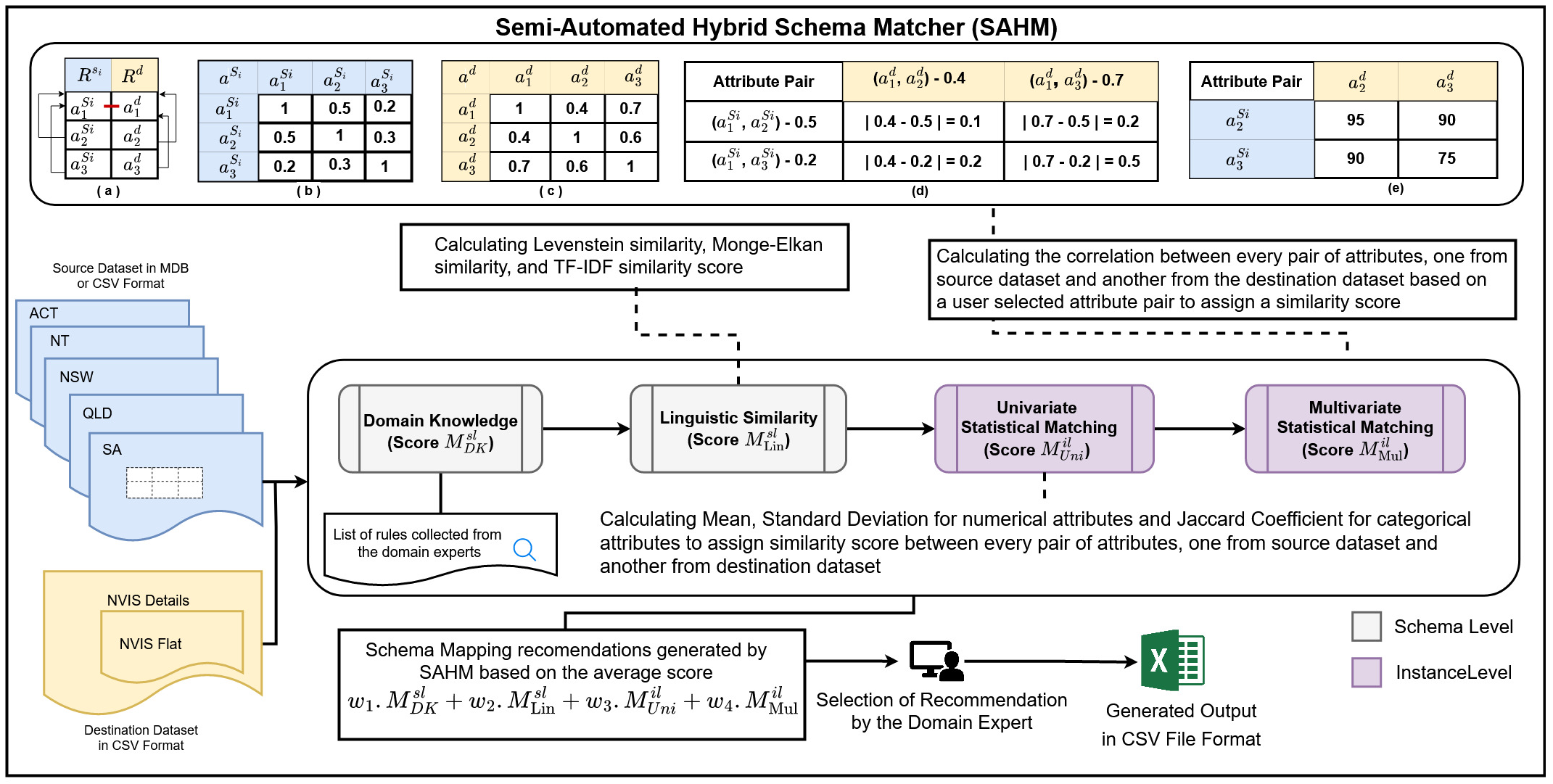} 
\caption{ System architecture of the proposed method  }
\label{fig:dist}
\end{figure*}

The National Vegetation Information System (NVIS)\citep{nvismanual} is an ongoing collaborative initiative between the Australian federal government, and state/territory governments. Here on in, we shall refer to both state and territory governments simply as state government. The NVIS integrates vegetation information such as structural and floristics data from the various state governments into a federally consolidated database. Figure \ref{fig:dist} illustrates this setting. In the figure, the NVIS data, also known as the destination data is denoted by $R^d(a^d_1,\cdots,a^d_{n_d})$. Each state data, also known as the source data is denoted by $R^{s_i}(a^{s_i}_1,\cdots,a^{s_i}_{n_{s_i}})$, where $i$ denotes a specific instance of a state.
\begin{figure}[th!]
    \centering
\includegraphics[scale=.4]{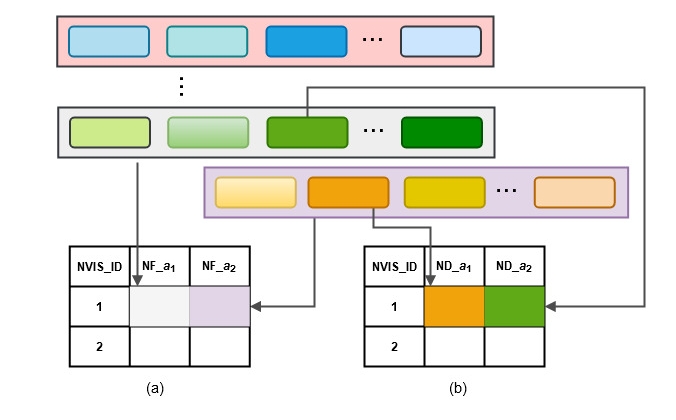}%
\caption{ Different schemes of NVIS database (a) NVIS Flat (NF) scheme  (b) NVIS Details (ND) scheme}
\label{fig:nvis_scheme}

\end{figure}
NVIS framework has been developed as a comprehensive data system to compile information on the extent and distribution of vegetation types from different states. The framework has two different databases representing two levels of abstraction, namely \emph{NVIS Details} (ND) and~\emph{NVIS Flat} (NF). ND provides detailed information on each record with attributes containing information on strata, species and structure. NF is a convenient representation of ND where each attribute (resp. attribute value) in NF is a combination of 2 or more attributes (resp. attribute values) from ND. Figure \ref{fig:nvis_scheme} illustrates these two databases.  For example, in the current NVIS scheme, an NF attribute value is 
``$U+ \hat{A}cacia sclerosperma,Acacia victoriae\backslash\hat{t}ree\backslash6\backslash i;M \hat{A}cacia eremaea,\\Acacia ramulosa var. linophylla,Senna sp.\backslash \hat{s}hrub \backslash 4 \backslash r;G \hat{C}enchrus ciliaris\\\backslash \hat{t}ussock grass \backslash 2 \backslash r$" 
delimited by ``;" represents strata, species and structural information respectively. Each value represents a separate piece of information For example ``U+" which stands for ``upper strata" in the NF attribute value above represents strata information. In the ND scheme however, this information will be recorded as a separate attribute value. 

Each state maintains their own vegetation database which may have a different schema compared to the NVIS schema. Further, the databases are often stored in different formats, for example, Australian Capital Territory (ACT), Northern Territory (NT), Western Australia (WA), and South Australia (SA) use a CSV format whereas Queensland (QLD) use an MDB format. The common practice is to transform all the different formats into one format manually prior to integrating the data into NVIS framework. At present, while integrating state datasets into NVIS framework, domain experts follow the rules stated in Australian Vegetation Attribute Manual (v7) (AVAM) \citep{nvismanual} for mapping data. For example, `Foliage cover' having values 70-100, 30-70, 10-30, $<$10, 0 and 0-5 can be represented as `Cover code' with values `d', `c', `i', `r', `bi' and `bc' respectively which is described in Table 7 in AVAM. Thus, the AVAM plays a vital role by providing guidelines for data integration in the manual process. This process is, however, complicated by inconsistencies in vegetation survey practices which may make data seemingly better aligned to one of the ND or NF schemas, albeit not completely. A main challenge of this integration process is to identify which NVIS schema is most aligned with the state data and further, to accurately determine matching attributes accross the databases.

\subsection{Related Work}
A number of schema matching techniques ~\citep{dumas,emd,distributionbased,semprop,embdi} and tools \citep{comapp1,cupid,coma,hm,fm,fa,am} have been developed in the field of data integration. CUPID \citep{cupid} is a hybrid approach to improve generic schema matching which exploits schema elements, schema structure, and linguistic-based matching to map schemas. COMA \citep{coma} is a generic schema matching approach, which combines simple, hybrid, and reuse-oriented matchers in a flexible way.  
COMA also represents a suitable environment for testing different algorithms and evaluating their individual or combined performances. 
Later they proposed an enhanced version, COMA++~\citep{comapp1,comapp2,comapp3} which supports a large number of matchers and requires user intervention for selecting the best combination of matchers. However, COMA++  reuses previous match results which require the same domain and entities for the new datasets. In addition, the action of matcher selection questions its feasibility if the user lacks proper knowledge about the system. DUMAS~\citep{dumas} is an instance-based schema matching algorithm that exploits the existence of duplicates among datasets for matching purposes. 
Distribution Based Matcher~\citep{distributionbased} is also an instance-based method that captures the relationships between different columns by comparing their respective data values. Distribution Based Matcher uses Earth Mover’s Distance (EMD) between pairs of columns for this purpose. To automate the analysis of protein crystallization, \citep{pc} proposed a linguistic-based schema matching technique that matches the schema elements based on their similarity values. However, considering only the linguistic similarity fails to analyze the element-level properties. EmbDI~\citep{embdi} is a framework that can facilitate schema matching on relational data by developing relational embeddings. EmbDI finds the relationships between the columns of two datasets by assessing their corresponding embeddings.  

A large number of tools have been implemented for schema matching based on various schema matching techniques. Harmony match engine \citep{hm} is a semi-automated schema matching tool that adds a linguistic pre-processing approach of textual documentation in conventional schema matching technique to speed up the findings of similarity across two schemas. To view and modify the identified similar schemas, this matcher adopts a score-based strategy that combines multiple match algorithms with a graphical user interface. FlexMatcher is another open-source tool \citep{fm} which uses a collection of supervised machine-learning techniques to train a schema matcher. In addition, Falcon-AO \citep{fa} and AgreementMaker \citep{am} are some other schema matching tools that support combined use of different linguistic,  structural and instance-based matching. 

\subsection{Challenges \& Opportunities}
There are 3 main issues in directly applying the existing techniques to our unique NVIS scenario. Firstly, the existing techniques do not provide a way to incorporate domain knowledge into the schema matching process. Indeed, existing techniques are often generalised and domain agnostic~\citep{pc}. In our NVIS scenario, we make the observation that available domain knowledge can be leveraged to improve schema matching tasks without increasing the dependence of the matching scheme on the existence of domain knowledge. 

Secondly, existing schema matching and mapping techniques~\citep{cupid,coma,distributionbased,semprop,embdi} have not fully explored the use of correlations among attributes in a source dataset $R^s$ as a means of mapping their attributes to the attributes in the destination dataset $R^d$. We term this novel approach \emph{Multivariate Statistical Matching}. The crux of this approach is based on the assumption that given any two attributes $a^s$ and $a^d$ found to be a match then if attribute $a^s_j$ is most correlated to $a^s$ and $a^d_k$ is most correlated to $a^d$, then it is likely that $a^s_j$ and $a^d_k$ are equivalent.

Thirdly, many existing systems expect a relational database and an ontology as input \citep{automatic_discovery_of_attribute,cupid}. However, this is not always the case. As described earlier, NVIS datasets may not strictly satisfy the tenets of a relational model (c.f. NF). At the same time the states do not provide any ontologies for their dataset schemas. Thus any proposed solution must be cognizant of this.

In the following, we describe how our proposed solution explores the opportunities described above to address the challenge of schema mapping in the NVIS scenario.

\section{Proposed Method} \label{proposed_method}
In this work, we propose a two-tiered ensemble schema matching decision support framework called \emph{SAHM} illustrated in Figure \ref{fig:dist}. The first tier function in SAHM, called \emph{schema level matcher} and denoted $M^{sl}(R^d, R^s) = score^{sl}$ extracts the schema information from $R^d$ and $R^s$ respectively and returns a list of matching scores $score^{sl}$ for each attribute pair $\langle a^s, a^d\rangle$. Similarly, the second tier function, called \emph{instance level matcher} and denoted $M^{il}(R^d, R^s) = score^{il}$ extracts information from the instances in $R^d$ and $R^s$ respectively and returns a list of matching scores $score^{il}$ for each attribute pair $\langle a^s, a^d\rangle$. Finally both scores, $score^{sl}$ and $score^{il}$ are aggregated to produce a final score denoted $score$ for each attribute pair $\langle a^s, a^d\rangle$. Thus, SAHM takes as an input the two databases and generates a set of tuples containing all attribute-pairs and their respective scores \emph{i.e.} $SAHM(R^d, R^s) = \{(\langle a^s, a^d\rangle,score)_i\}$.

We note that, each matching function $M^{sl}$ (\emph{resp.} $M^{il}$) is an ensemble of multiple matching functions. These functions are detailed as follows.

\subsection{Schema Level Matching ($M^{sl}$)}
Schema level matching uses information from the attribute names to estimate the matching score between any two pairs of attributes $\langle a^s, a^d\rangle$ from the source and destination databases respectively. 
Table \ref{tab_exall} provides an example entry of a source and destination table from state and NVIS databases respectively. This example will be considered throughout the paper for illustrating different matchers. Schema level matching is an ensemble that comprises of two schema level based matchers as follows.

\begin{table}[htbp]
  \small
    \centering
  \caption{An example entry from the state and NVIS databases}
  \label{tab_exall}
  \begin{subtable}{.45\linewidth}
  \smaller
  \centering
  \caption{Source database (state database)}
  \begin{tabular}{@{}cc@{}}
  \toprule
\multicolumn{1}{c}{u\_heightCode} & \multicolumn{1}{c}{treesp\_3} \\ \midrule
8 & Eucalyptus rossii \\
0 & Eucalyptus bridgesiana \\
2 & Allocasuarina verticillata \\ \bottomrule
\end{tabular}%
    \hspace{.5cm}%
  \end{subtable}\hfill
  \begin{subtable}{.45\linewidth}
  \smaller
  \centering
  \caption{Destination database (NVIS database)}
      \begin{tabular}{@{}cc@{}}
      \toprule
\multicolumn{1}{c}{u\_height\_class} & \multicolumn{1}{c}{u\_species\_3} \\ \midrule
0 & Eucalyptus bridgesiana \\
1 & Atalaya hemiglauca \\
5 & Pomaderris aspera \\ \bottomrule
\end{tabular}%
    \hspace{.5cm}%
 \end{subtable}
\end{table}

\textbf{Domain Knowledge ($M^{sl}_{DK}$):} 
The first schema level matcher employs domain knowledge, where it is available. In our NVIS scenario, we leverage the wealth of knowledge accumulated by the domain experts, and recorded in the NVIS manual~\citep{nvismanual}. These are encoded as \emph{if-then-} rules and fed into the SAHM system. The domain knowledge based schema level matcher denoted $M^{sl}_{DK}$ iterates through the rules to identify if the pair $\langle a^s, a^d\rangle$ corresponds to a valid rule. That is, given a source and destination dataset $R^s$ and $R^d$, and the set of attribute pairs $\{\langle a^s, a^d\rangle\}^{n_s\times n_d}$, the domain knowledge schema level matcher for each pair $\langle a^s, a^d\rangle$ is calculated as follows : 
\begin{equation}
\label{eqdomk}
    M^{sl}_{DK}(\langle a^s, a^d\rangle) =
    \begin{cases}
    1 & \parbox[t]{0.6\linewidth}{if ``if $a^s$ (\emph{resp.} $a^d$) then $a^d$ (\emph{resp.} $a^s$)'' is valid; }\\
    0 & \text{otherwise}
    \end{cases}
\end{equation}

Although domain knowledge can play a vital role in schema matching, it is not a necessary requirement for SAHM. When domain knowledge is not available, $M^{sl}_{DK}$ can be excluded from the ensemble $M^{sl}$ and still yield reasonably good results as demonstrated in the experiments section. In this case, $M^{sl}_{DK}(\langle a^s, a^d\rangle) = 0$ for all pairs $\langle a^s, a^d\rangle$.  

\textbf{Linguistic based Matching $(M^{sl}_{Lin})$ :} The second schema level matcher, employs 3 linguistic based metrics to identify a match.  The metrics are Levenshtein similarity~\citep{levenshtein}, Monge-Elkan similarity~\citep{monge} and TF-IDF~\citep{tfidf}. Levenshtein similarity calculates the similarity between two strings based on edit distance. Levenshtein similarity is in the range of $[0,1]$, when $a^s = a^d$ the Levenshtein similarity is $1$. In particular, Levenshtein similarity, $Sim_{LEV}$, between two string $a^s$ and $a^d$ is calculated using Equation \eqref{eqsimlev}. Here, $|a^s|$ and $|a^d|$ denote the number of letters in $a^s$ and $a^d$, respectively. Therefore, max($|a^s|$, $|a^d|$) denotes the number of letters of the longer string out of $a^s$ and $a^d$.
\begin{equation}
\label{eqsimlev}
    Sim_{\mathrm{LEV}}(a^s,a^d) =  1 - \frac{\mathrm{editDistance}(a^s,a^d)}{\max(|a^s|,|a^d|)}
\end{equation}

For example, the Levenshtein distance between `$u\_heightCode$' and `$u\_height\newline \_class$' from Table \ref{tab_exall} is 5, which consists of insertion of two letters (i.e. `$\_$', `$s$') at positions 8 and 12 respectively and three replacements (i.e., $`o\text{'}\rightarrow `l\text{'}$, $`d\text{'}\rightarrow `a\text{'}$ and $`e\text{'} \rightarrow `s\text{'}$) at positions 10, 11 and 12 respectively of `$u\_heightCode$'. The similarity score calculated using equation \eqref{eqsimlev}  will be, $1- (5 /14) = 0.64$.

As previously seen in Figure~\ref{fig:nvis_scheme}, some attributes may be complex compound names consisting of multiple words. This is peculiar to the NVIS environment since attributes names may consist of the plant genealogy. We adapt Monge-Elkan and TFIDF techniques for such a case. In Monge-Elkan similarity, an average Levenshtein similarity is calculated at the token level for all tokens. Monge-Elkan similarity is in the range of $[0,1]$. Given string $a^s$ (resp. $a^d$), let $T(a^s)$ (resp. $T(a^d)$) be the tokens in $a^s$ (resp. $a^d$). The Monge-Elkan similarity, $Sim_{ME}$, is computed following Equation \eqref{eqsimmonge}. In Equation \eqref{eqsimmonge}, $Sim_{LEV}$($T(a^s)$,$T(a^d)$) is the Levenshtein similarity of $T(a^s)_i$ (i.e. the $i$th token in $a^s$) and $T(a^d)_j$ ( i.e. the $j$th token in $a^d$) (\emph{cf.} Equation \eqref{eqsimlev}).
\begin{equation}
    \label{eqsimmonge}
    Sim_{ME}(a^s,a^d) = \frac{1}{|T(a^s)|}\sum^{T(a^s)}_{i=1}\max\{Sim_{LEV}(T(a^s)_i,T(a^d)_j)\}^{T(a^d)}_{j=1}
\end{equation}

For example, let us consider two attributes, `$treesp\_3$' and `$u\_species\_3$' from Table \ref{tab_exall}. The inner function is used to compute the scores of the best matching token. Thus, $Sim_{LEV}$(`treesp',`u') $= 0$, $Sim_{LEV}$(`treesp',`species') $= 0.28$, $Sim_{LEV}$(`treesp',`3') $= 0$, $Sim_{LEV}$(`3',`u') $= 0$, $Sim_{LEV}$(`3',`species') $= 0$, $Sim_{LEV}$(`3',`3') $= 1$. Finally, Monge\-Elkan similarity score calculated using Equation \eqref{eqsimmonge} will be $1/2*(0.28+1) = 0.64$.

In the case of TF-IDF \citep{tfidf}, each attribute name in the source and destination databases is tokenized. Next, each attribute name is treated as a document and each token from the attribute is treated as a term. TF-IDF score is calculated for each token. The TF-IDF similarity score between any two source and destination attribute names $a^s$ and $a^d$  denoted $Sim_{TFIDF}(a^s, a^d)$ is calculated using equation \eqref{tfidf}. 

\begin{align}
\label{tfidf}   
Sim_{TFIDF}&( a^s, a^d) = \frac{\prod\limits_{\forall a \in \{a^s,a^d\}}\{\sum\limits_{t \in \{T(a^s) \cap T(a^d) \}}TFIDF(t,a)\}}{\prod\limits_{\forall a \in \{a^s,a^d\}}\{\sum\limits_{t \in \{T(a^s) \cup T(a^d) \}}TFIDF(t,a)\}}
\end{align}

Here, we use the symbol $a^s \cap a^d$ to mean the common tokens between the attributes and $a^s \cup a^d$ to mean the union of the tokens. The function $TFIDF(t, a)$ calculates the TF-IDF score for the token $t$ of a given attribute $a$. The score is normalised to the range $[0,1]$. For example, the TF-IDF similarity score between source attribute ($a^s$) `$u\_heightCode$' and destination attribute ($a^d$) `$u\_height\_class$' (\emph{cf.} Table \ref{tab_exall}) is calculated as follows:
\small{ 
\begin{multline*}
{\displaystyle
Sim_{TFIDF}(u\_heightCode, u\_height\_class)   =} \\ 
{\displaystyle
\dfrac{ \prod\limits_{\forall a \in \{u\_heightCode, u\_height\_class\}}\{\sum\limits_{t \in \{u, height \}}TFIDF(t,a)\}}{\prod\limits_{\forall a \in \{u\_heightCode, u\_height\_class\}}\{\sum\limits_{t \in \{u, height, Code, class\}}TFIDF(t,a)\}}
}\\ \\
{ =\frac{\left(
\splitdfrac{(TFIDF(u,u\_heightCode)+TFIDF(height,u\_heightCode))}{\times(TFIDF(u,u\_height\_class)+TFIDF(height,u\_height\_class))}\right)}{\left(
\splitdfrac{\splitdfrac{(TFIDF(u,u\_heightCode)+TFIDF(height,u\_heightCode)}{+TFIDF(Code,u\_heightCode)+TFIDF(class,u\_heightCode))}}{\times \splitdfrac{(TFIDF(u,u\_height\_class)+TFIDF(height,u\_height\_class)}{+TFIDF(class,u\_height\_class)+TFIDF(Code,u\_height\_class))}}\right)}
}\\ \\
=\dfrac{(0.04+0.10)\times(0.04+0.10)}{(0.04+0.10+0.20+0)\times(0.04+0.10+0.20+0)} =\dfrac{0.0196}{0.1156} = 0.17 
\end{multline*}
}

Similarly, the TF-IDF similarity score, $Sim_{TFIDF}(u\_heightCode, u\_species3)$ (\emph{cf.} Table \ref{tab_exall}), is 0.013 which is smaller than the score of $Sim_{TFIDF}(u\_heightCode, u\_\newline height\_class) = 0.17$ due to the former having more dissimilar tokens.

After calculating individual linguistic scores, a weighted sum is taken to generate the final Linguistic similarity score for any pair $\langle a^s, a^d\rangle$. That is, given a source and destination dataset $R^s$ and $R^d$, and the set of attribute pairs $\{\langle a^s, a^d\rangle\}^{n_s\times n_d}$, the linguistic schema level matcher for each pair $\langle a^s, a^d\rangle$ is calculated as follows:  
\begin{multline}
\label{eqsimlin}
  M^{sl}_{Lin}(\langle a^s, a^d\rangle)= g_1\cdot Sim_{Lev}(\langle a^s, a^d\rangle) + g_2\cdot Sim_{ME}(\langle a^s, a^d\rangle)\\ + g_3\cdot Sim_{TFIDF}(\langle a^s, a^d\rangle)    
\end{multline}
  
where $g_1+g_2+g_3 = 1$. 

Note that the number and types of linguistic metrics used may be considered as a user defined parameter. However, our experiments show that our configuration yields comparatively more accurate results for our application.

\subsection{Instance Level Matching ($M^{il}$)}
Instance level matching uses information from the attribute values of any two attributes from the source and destination databases $\langle a^s, a^d\rangle$ to estimate the matching score between them. The ensemble comprises of 2 instance level based matchers as follows.

\textbf{Univariate Statistical Matching $(M^{il}_{Uni})$ :} This matcher computes the the similarity between a pair of attributes $\langle a^s, a^d\rangle$ using the attribute values. There are two variants of this matcher, one for categorical attributes and another for numerical attributes. In numerical attributes, we make use of the  mean and standard deviation of the attribute, while in categorical attributes, we make use of the Jaccard coefficient. Specifically, in the case of numerical attributes, the mean identifies the central tendencies of the two attributes, whose closeness can be measured. Further, their standard deviations can be used to measure the closeness of their dispersions from their respective central tendencies. We adopt a weighted sum of the normalised absolute differences in the mean and standard deviation to calculate univariate score.  For two categorical attributes, we adopt the Jaccard Coefficient \citep{statisticalsm}. Jaccard coefficient ensures that when the two attributes have a higher number of attribute values in common, then they are considered to be more similar. Thus, given a source and destination dataset $R^s$ and $R^d$, and the set of attribute pairs $\{\langle a^s, a^d\rangle\}^{n_s\times n_d}$, the univariate statistical instance level matcher for each pair $\langle a^s, a^d\rangle$ is calculated as follows:

\begin{equation}
    M^{il}_{Uni}(\langle a^s, a^d\rangle) =
    \begin{cases}
    1 - (\frac{0.8 \cdot \lvert{\mu(a^s)-\mu(a^d)}\rvert}{\max(\mu(a^s), \mu(a^d))} + \frac{0.2 \cdot \lvert{SD(a^s)-SD(a^d)}\rvert}{\max(SD(a^s), SD(a^d))}) & \parbox[t]{0.6\linewidth}{if $a^s$, $a^d$ \\are numerical }\\\\
    \frac{|\{T(a^s)\cap T(a^d)\}|}{|\{T(a^s)\cup T(a^d)\}|} & \text{otherwise}
    \end{cases}
    \label{eal}
\end{equation}

In equation \eqref{eal}, $\mu$ and $SD$ are the mean and standard deviation respectively. $M^{il}_{Uni}(\langle a^s,a^d \rangle) \in [0,1]$. Note that when the attributes are numerical and their means and standard deviations approach each other, the score approaches 1 \emph{i.e} $M^{il}_{Uni}(\langle a^s, a^d\rangle) \rightarrow 1 $. Similarly when the attributes are categorical and the number of common attribute values approach each other then the score also approaches 1 \emph{i.e} $M^{il}_{Uni}(\langle a^s, a^d\rangle) \rightarrow 1 $. We assigned larger weight on mean value in equation~\ref{eal} as it is more significant in contributing similarity, based on our experimental results. 
If one of the two attributes is numerical and the other attribute is categorical, then the numerical attribute is first converted into a categorical attribute through discretization \citep{discretization}, and then the Jaccard Coefficient is computed. 

For example, in Table \ref{tab_exall}, there are both numerical and categorical attributes. Thus, the univariate score of attribute pair $\langle u\_heightCode, u\_height\_class \rangle$ is,
\small{\begin{align*}
     M^{il}_{Uni}(\langle u\_heightCode, u\_height\_class\rangle) \\\notag
     &= 1-(\frac{0.8 \cdot \lvert{\mu(u\_heightCode)-\mu(u\_height\_class)}\rvert}{\max(\mu(u\_heightCode), \mu(u\_height\_class))} \\\nonumber
     & + \frac{0.2 \cdot \lvert{SD(u\_heightCode)-SD(u\_height\_class)}\rvert}{\max(SD(u\_heightCode), SD(u\_height\_class))})\\\nonumber
     &= 1-(\frac{0.8 \cdot \lvert{3.33-2}\rvert}{\max(3.33,2)}+\frac{0.2 \cdot \lvert{3.39-2.16}\rvert}{\max(3.39,2.16)})\\ \nonumber
     &= 0.61
\end{align*}}
Again, the univariate score of attribute pair $\langle treesp3, u\_species3 \rangle$ is
\[
\begin{split}
      M^{il}_{Uni}(\langle treesp3, u\_species3\rangle) \\\notag
     &=\frac{|\{T(treesp3)\cap T(u\_species3)\}|}{|\{T(treesp3)\cup T(u\_species3)\}|} \\ \nonumber
     &= \frac{1}{5}\\ \nonumber
     &= 0.2
\end{split}\]

\textbf{Multivariate Statistical Matching $(M^{il}_{Mul})$ :}  This novel matcher is based on 
the correlation between source and destination attribute pairs $\langle a^s, a^d \rangle$. The intuition behind this approach is that given a pair $\langle a^s, a^d \rangle$ that are known to be a match, the correlation between $a^s$ and other attributes in the source database $R^s$ will present similar characteristics as the correlation between $a^d$ and other attributes in the destination database $R^d$. For the given attribute pair $\langle a^s,a^d \rangle$, known to be a match, the multivariate statistical matching score $M^{il}_{Mul}$ generates the Pearson's Correlation between the source attribute $a^s$ and other attributes $a^s_j$ from source dataset $R^S$. A ranked list of attributes $[(a^s_q, pc_{a^s_q}) \preceq \cdots \preceq (a^s_r, pc_{a^s_r})]$ according to the Pearson's correlation score is generated. Similarly, the multivariate statistical matching score $M^{il}_{Mul}$ generates the Pearson's Correlation between the destination attribute $a^d$ and other attributes $a^d_k$ from destination dataset $R^d$. A ranked list of attributes $[(a^d_q, pc_{a^d_q}) \preceq \cdots \preceq (a^d_r, pc_{a^d_r})]$ according to the Pearson's correlation score is also generated. For every pairwise combination of attributes in $[(a^s_q, pc_{a^s_q}) \preceq \cdots \preceq (a^s_r, pc_{a^s_r})]$ and $[(a^d_q, pc_{a^d_q}) \preceq \cdots \preceq (a^d_r, pc_{a^d_r})]$, that is for every pair $<a^s_j, a^d_k>$ the absolute distance $|pc_{a^s_j} - pc_{a^d_k}|$ in the Pearson's Correlation values is calculated. 
Thus for any given pair $\langle a^s, a^d \rangle$ that are known to be a match, a set of distance values $\{|pc_{a^s_j} - pc_{a^d_k}|\}^{(n_d-1) \times (n_s -1)}$ is generated, where $n_s$ (resp. $n_d$) is the number of attributes in the source (resp. destination) database. The distance values are normalised by dividing by the maximum distance $2$ and inverted by subtracting from $1$. Thus, for a given $\langle a^s, a^d \rangle$, $M^{il}_{Mul}(\langle a^s, a^d \rangle)$ generates a set of similarity scores  $ M^{il}_{Mul}(\langle a^s, a^d \rangle) =\{Sim_{pc}(a^s_j,a^d_k)\}^{(n_d-1) \times (n_s -1)}$, where $Sim_{pc}(a^s_j,a^d_k) = 1 - \frac{|pc_{a^s_j} - pc_{a^d_k}|}{2}$. Thus for all previously known matching pairs $\{\langle a^s, a^d \rangle\}$, $M^{il}_{Mul}(\langle a^s, a^d \rangle)$ can be calculated for each pair $\langle a^s, a^d \rangle$ and the corresponding scores $Sim_{pc}(a^s_j,a^d_k)$ generated for each previously known matching pair $\langle a^s, a^d \rangle$ averaged to obtain the final score $\bar{Sim_{pc}}(a^s_j,a^d_k)$.

\begin{figure*}[th!]
\centering
\subfloat[\label{fig:cLL1}]{%
  \includegraphics[height=2.3cm,width=.3\linewidth]{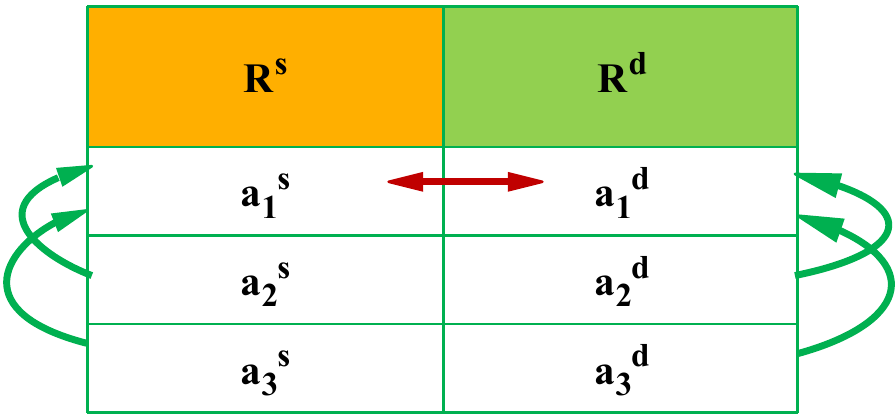}%
  \label{fig:ex1}
}\hfil
\subfloat[\label{fig:rCll1}]{%
  \includegraphics[height=2.3cm,width=.29\linewidth]{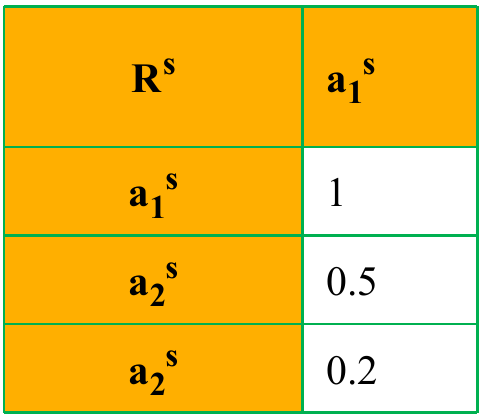} %
  \label{fig:ex2}
}\hfil
\subfloat[\label{fig:cLL2}]{%
  \includegraphics[height=2.3cm,width=.29\linewidth]{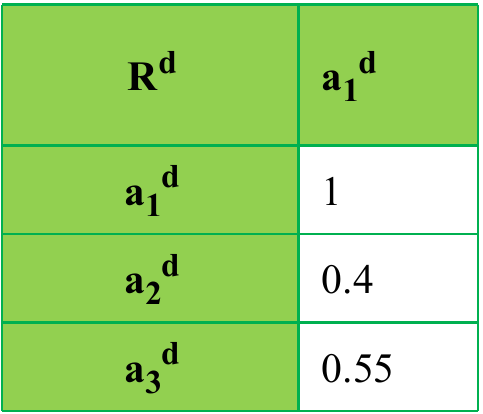}%
  \label{fig:ex3}
}\hfil
\subfloat[\label{fig:rCll2}]{%
  \includegraphics[height=2.2cm,width=0.62\linewidth]{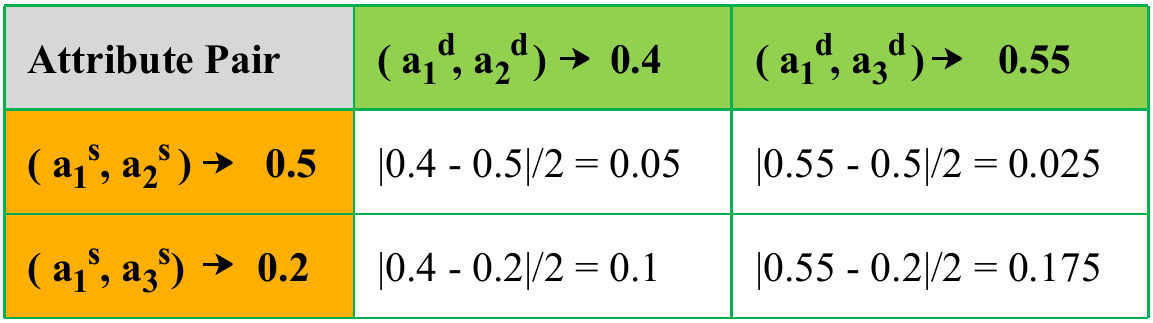} %
  \label{fig:ex4}
}\hfil
\subfloat[\label{fig:rCll}]{%
  \includegraphics[height=2.2cm,width=0.31\linewidth]{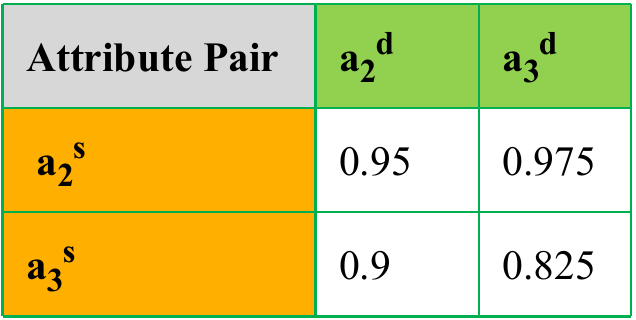} %
  \label{fig:ex5}
}\hfil

\caption{ Example of Normalised Multivariate Scoring. (a) Source-Destination attribute list where user-defined attribute pair is ($a_{1}^s \leftrightarrow a_{1}^d$) (b) Source table with correlation scores (c) Destination table with correlation scores (d) Calculation of distances for source-destination pairs (e) Multivariate matching scores}
\label{fig:ex}
\end{figure*}

Figure \ref{fig:ex} depicts an example scenario of multivariate scoring scheme where \{$a_1^s, a_2^s, a_3^s$\} and \{$a_1^d, a_2^d, a_3^d$\} are attributes from source and destination dataset respectively (Figure \ref{fig:ex1}). 
In the proposed normalised multivariate scoring scheme, at first, user input is required where user defines a known match for a specific source-destination pair. In the example in Figure \ref{fig:ex1}, this user-defined pair ($a_1^s, a_1^d$) is indicated by red arrow. As the relationship between one pair is already defined, the proposed multivariate scoring scheme will suggest destination attributes for the other two source attributes. The multivariate scores between source attributes are presented in Figure \ref{fig:ex2} which indicate the relationship between source-source attribute pair based on the user-defined relationship. The higher value indicates a stronger relationship.
Similarly, Figure \ref{fig:ex3} represents the multivariate scores between destination-destination attribute pairs. For every combination of source-destination pair, Figure \ref{fig:ex4} exhibits the normalised absolute distance value of multivariate scores found from source and destination attribute tables. Since Figure \ref{fig:ex4} shows the normalised absolute distance between source-destination pair, the lowest score indicates the most correlated pair. Finally, the multivariate score (Figure \ref{fig:ex5}) is calculated by subtracting each of the scores found in previous step from 1. 

\begin{algorithm}[htp]
  \caption{SAHM Scoring Algorithm}
 \label{algoM}
  \scriptsize
  \SetAlgoNoEnd  \DontPrintSemicolon
  \SetKwFunction{algo}{SAHM}
  \SetKwProg{myalg}{Algorithm}{}{}
  \myalg{\algo{}} {
   \nl \ForEach{$a^d \longleftarrow  R^d$}
  {
    \nl \ForEach{$a^s \longleftarrow  R^s$}
    {
        \nl  $score^{sl}_{DK}$, $score^{sl}_{Lin}$ $\longleftarrow$ \mbox{SLM}($a^s,a^d$) \;
        
        \nl  $score^{il}_{Uni}$, $score^{il}_{Mul}$ $\longleftarrow$ \mbox{ILM}($a^s,a^d$) \;
        \nl  $score(a^s,a^d)$ $\longleftarrow$ $w_1\cdot score^{sl}_{DK}+w_2\cdot score^{sl}_{Lin}+w_3\cdot score^{il}_{Uni} +w_4\cdot score^{il}_{Mul}$ \;
    
    }
	    
     }
  
  \nl \KwRet $score(a^s,a^d)$\;}{}

  \setcounter{AlgoLine}{0}
  \SetKwFunction{proc}{SLM}
  \SetKwProg{myproc}{Procedure}{ : $M^{sl}_{DK},M^{sl}_{Lin}$}{}
  \myproc{\proc{$a^s$, $a^d$}}{
    
    \nl \mbox{Calculate} $M^{sl}_{DK}$ \mbox{ using equation \eqref{eqdomk}}\;
    \nl \mbox{Calculate} $M^{sl}_{Lin}$ \mbox{ using equation \eqref{eqsimlin}}\;
    
  \nl \KwRet $M^{sl}_{DK},M^{il}_{Lin}$ \;
  }{}

  \setcounter{AlgoLine}{0}
  \SetKwFunction{proc}{ILM}
  \SetKwProg{myproc}{Procedure}{ : $M^{il}_{Uni},M^{il}_{Mul}$}{}
  \myproc{\proc{$a^s$, $a^d$}}{
    
    \nl \mbox{Calculate} $M^{il}_{Uni}$ \mbox{ using equation \eqref{eal}}\;
    
    \nl \ForEach{$\langle a^s, a^d \rangle$ $\longleftarrow$  Known Matching Pairs  $\{\langle a^s, a^d \rangle\}$} 
    {
	    \nl  $[(a^s_q, pc_{a^s_q}) \preceq \cdots \preceq (a^s_r, pc_{a^s_r})]$ $\longleftarrow$ \mbox{CalcCorrelation}($a^s,a^s_j$) \;
        
        \nl  $[(a^d_q, pc_{a^d_q}) \preceq \cdots \preceq (a^d_r, pc_{a^d_r})]$ $\longleftarrow$ \mbox{CalcCorrelation}($a^d,a^d_k$) \;
        
        \nl  $CR_{v}$ $\longleftarrow$ \mbox{CalcMirrorCorrelation}($[(a^s_q, pc_{a^s_q}) \preceq \cdots \preceq (a^s_r, pc_{a^s_r})] , [(a^d_q, pc_{a^d_q}) \preceq \cdots \preceq (a^d_r, pc_{a^d_r})]$)\;
        
        \nl  Add $CR_{v}$ into $M^{il}_{Mul}$ \;
     }
  \nl\KwRet $M^{il}_{Uni},M^{il}_{Mul}$ \;
  }{}

  \setcounter{AlgoLine}{0}
  \SetKwFunction{proc}{CalcCorrelation }
  \SetKwProg{myproc}{Procedure}{ : $CR_{*}$}{}
  \myproc{\proc{$a^*$, $a^*_*$}}{
  \nl $CR_{*}$ = [] \; 
  \nl \ForEach{$a$ $\longleftarrow$  $a^*_*$}
  {
      \nl $\langle a^*_q, pc_{a^*_q}\rangle$  $\longleftarrow$ \mbox {PearsonCorrelation}($a^*,a$) \;
      
      
      \nl Add $\langle a^*_q, pc_{a^*_q}\rangle$ to $CR_{*}$ \;

  }
  \nl \KwRet $CR_{*}$ \;
  }{}

  \setcounter{AlgoLine}{0}
  \SetKwProg{myproc}{Procedure}{ : $CR_L$}{}
  \SetKwFunction{proc}{CalcMirrorCorrelation}
  \myproc{\proc{$CR_s$,$CR_d$}}{
  
  \nl \ForEach{$a^s_j,pc_{a^s_j}$ $\longleftarrow$  $CR_s$}
  {
      \nl \ForEach{$a^d_k,pc_{a^d_k}$ $\longleftarrow$  $CR_d$}
      {
        \nl $Dist$ $\longleftarrow$  $\|pc_{a^s_j} - pc_{a^d_k}\|$ \;
        
        \nl $Sim_{pc}(a^s_j,a^d_k)$ $\longleftarrow$ $1 - ($Dist$/2)$ \;
        
        \nl  \eIf{ $Sim_{pc}(a^s_j,a^d_k)$ not exist in $CR_L$  }
        {
            \nl Add $Sim_{pc}(a^s_j,a^d_k)$ in $CR_L$ 
        }
        {
           \nl Update $CR_L$ with average of existing $Sim_{pc}(a^s_j,a^d_k)$ and new $Sim_{pc}(a^s_j,a^d_k)$ \;
        }
      }
  }
  \nl \KwRet $CR_L$\;
  }{}
  \end{algorithm} 

 As shown in Algorithm~\ref{algoM}, a weighted sum $w_1\cdot M_1^{sl}+w_2\cdot M^{sl}_{Lin}+w_3\cdot M^{il}_{Uni}+w_4\cdot M^{il}_{Mul}$ is used to generate the final matching scores for each pair of source and destination attribute. Where $w_1+w_2+w_3+w_4 = 1$.  In this scheme, we have assigned equal values to the weights however, based on the context of the problem, preference can be made to different matchers of SAHM. 

In this section, a novel two tiered schema matching technique based on the database schema and instances has been introduced. In the following section, we demonstrate the effectiveness of our novel approach SAHM.

\label{sec-evaluation}


%

\section{Experiments and Results}
This section presents the performance evaluation of the proposed method SAHM in comparison with existing well known schema matching and mapping techniques listed in the valentine framework~\citep{valentine}. The techniques include COMA, CUPID, Similarity flooding, Jaccard Leven Matcher, Distribution Based Matcher, and EmbDI. In our evaluation, we consider effectiveness measured by F1 and accuracy scores, as well as the efficiency measured by runtime values. Further, an ablation study is conducted to investigate the contributions of each matcher in the proposed method SAHM. 
 

\begin{table}[h]
	\small
\centering
\caption{Proprietary NVIS Datasets}
\label{tab:spec}
\begin{tabular}{|c|c|c|}
\hline
\textbf{Dataset} & \textbf{\# Attributes} & \textbf{\# Records} \\\hline
{ACT} & 35 & 19226  \\ \hline
{NT} & 40 &  588 \\ \hline
{WA} & 17 &  2158 \\ \hline
{SA} & 20 &  1167 \\\hline
{QLD} & 20 &  14610 \\ \hline
{NVIS} & 68 &  10775 \\ \hline
\end{tabular}%
\end{table}

\begin{table}[H]
	\small
	\centering
	\caption{Musicians Datasets}
	\label{tab:spec_mus}
	\begin{tabular}{|p{1.7cm}|p{2cm}|p{1.6cm}|p{5cm}|}
		\hline
		
		\textbf{Dataset} & \textbf{\#Attributes} & \textbf{\#Records} & \textbf{Description} \\\hline
		{Unionable} &  20 & 10846 & A pair of source and destination datasets with the same set of attributes which may be verbatim or noised. It also contains some common instances which may be verbatim or noised  \\ \hline
		{View Unionable} &  13 & 5423& A pair of source and destination datasets which has no common instances but contains some common attributes which may be verbatim or noised. \\ \hline
		{Joinable} & 13 & 10846 & A pair of source and destination datasets which has some common instances and some common attributes. The common attributes may be verbatim or noised.   \\ \hline
		{Semantically Joinable} & 14 & 10846 &A pair of source and destination datasets which has some common instances and some common attributes. The common instances and attributes may be verbatim or noised.  \\\hline
		
	\end{tabular}%
\end{table}

\normalfont

{\bf Experimental setting:} All experiments were conducted on a Windows 10 Operating System with an AMD Ryzen 7 (PRO 3700Uw) 2.30GHz processor and 32 GB memory.
The performance of the proposed method has been evaluated on both the NVIS dataset and the Musicians dataset. The NVIS dataset involves proprietary state datasets from Western Australia (WA), South Australia (SA), Queensland (QLD), Northern Territory (NT) and Australian Capital Territory (ACT). Table \ref{tab:spec} shows the details of these datasets. Each state dataset is a source dataset to be matched with the NVIS dataset (destination dataset).

The Musicians dataset~\footnote{Available at https://bit.ly/3Kk3ryU} is a publicly available dataset of  American musicians generated by~\citep{valentine} from WikiData~\footnote{https://www.wikidata.org/wiki/Wikidata}. There are 4 variants of the musicians dataset and these are described in Table~\ref{tab:spec_mus}.

In all our experiments, the evaluation is based on how well the source dataset is matched to the destination dataset. It is worth noting that, the output of our proposed method SAHM leads to a ranking of the attributes of the source dataset to every attribute in the destination dataset. To calculate the accuracy, we consider the top N ranked attributes. We denote the top N ranked attributes by {\it Top\_*}, where * is the parameter on the number of attributes considered. By default, we always make use of \emph{Top\_1}, however, the user can change this parameter to obtain a higher number of ranked suggestions. For example, we demonstrate the performance when the top 1 to 4 ranked attributes (\emph{i.e.} \emph{Top\_1, Top\_2, Top\_3 \& Top\_4}), according to SAHM, are used in Figure \ref{fig:figb}. Further, we do not consider the domain knowledge matcher of SAHM in all experiments since it is trivial to show that the domain knowledge matcher aligns with the ground truth matches.

\subsection{Effectiveness of SAHM}

\begin{figure}[t]

\subfloat[\label{fig:figa}]{%
  \includegraphics[width=\linewidth, height =1.7in]{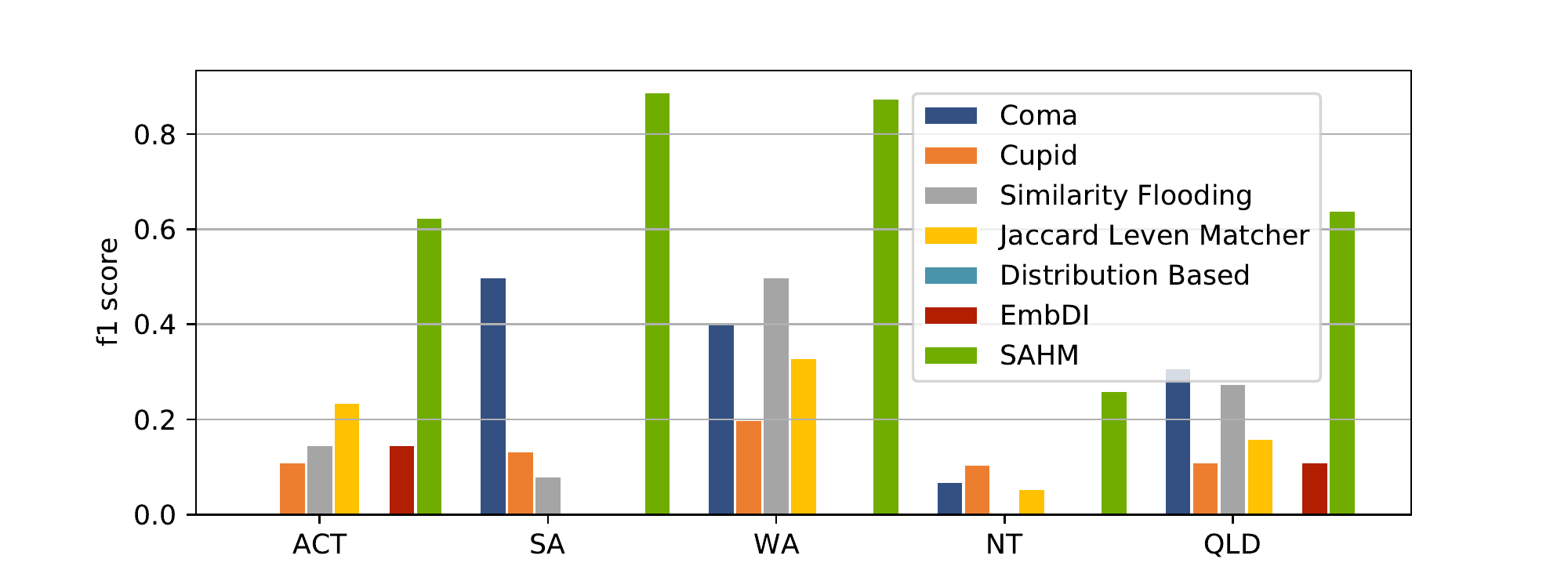}}
  
\subfloat[\label{fig:fig_mus}]{%
  \includegraphics[width=\linewidth,height =1.7in]{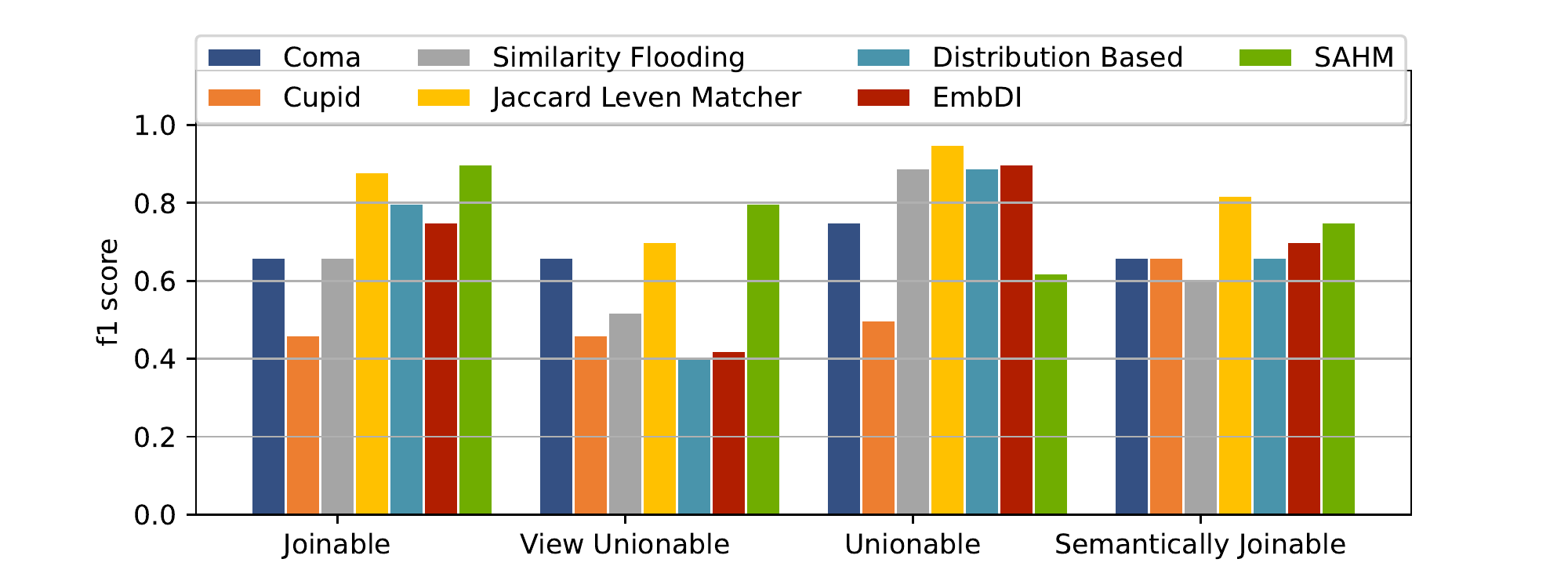}}%

\caption{F1 score comparison among different schema matching algorithms (a) Proprietary NVIS Datasets (b) Musicians Datasets} 
 \label{fig:results}
\end{figure}

In this experiment, we compared SAHM's effectiveness  with five other state-of-the-art schema matching algorithms by calculating the F1 scores. 
In this comparison, we have used the standard experiment setup of the valentine framework which returns the best possible matches between the state and NVIS dataset for COMA, CUPID, Similarity flooding, Jaccard Leven Matcher, Distribution Based Matcher, and EmbDI. In SAHM we consider the \emph{Top\_1} as our best match.

Figure \ref{fig:figa} is the results of our experiment on the proprietary datasets. It is observed that the overall F1 score of our proposed method outperforms the other methods across all the datasets. Overall, the NT state dataset performance was lower than the other states due to an unusual naming pattern of the attributes. It is also worth noting that the results indicate that the existing approaches are not always suitable. That is, the best match predicted by the existing techniques do not always align with the ground truth. For example, it can be seen that COMA is not able to correctly determine any true matches from the NVIS dataset within the ACT dataset while Jaccard Leven Matcher does not find any matches in the SA dataset. Further, EmbDI is not able to identify any matches in SA, WA and NT datasets. Distribution based approach is the worst performing technique as it does not find any true matches in all the datasets. These results not only show the effectiveness of SAHM but also its reliability across different state datasets. 

Figure~\ref{fig:fig_mus} is the results obtained from the publicly available musicians dataset. We observe that SAHM is very competitive and has the best performance in two out of the four datasets. In particular we notice that SAHM works best with the musicians datasets that has smaller number of attributes. SAHM has the best performance in Joinable and View Unionable datasets both having 13 attributes,  and performs second best in Semantically Joinable dataset which has 14 attributes. However, it is not competitive in the Unionable dataset which has 20 attributes. Recall from Table~\ref{tab:spec_mus} that the Unionable dataset has the same set of attributes in both the source and destination dataset pair, however the attributes in the source dataset are noised. In this scenario, SAHM does not effectively discriminate between the attributes. This is because of how the noise is generated. Quite often, the word ``Label" is added as noise to the attributes in the source dataset (\emph{e.g.} ``cityLabel"). In this case the schema level matchers will often mismatch the attributes in the destination dataset (\emph{e.g.} ``religionLabel"). We observe that in such a case, when the weighting scheme for SAHM is adjusted in favour of the instance based matching component, SAHM performs considerable better. In the future, we will explore how this weighting can be optimised for any given dataset.

\subsection{Efficiency of SAHM}
\begin{figure}[t]
	
	\subfloat[\label{fig:runtime_avg}]{%
		\includegraphics[width=\linewidth, height =1.7in]{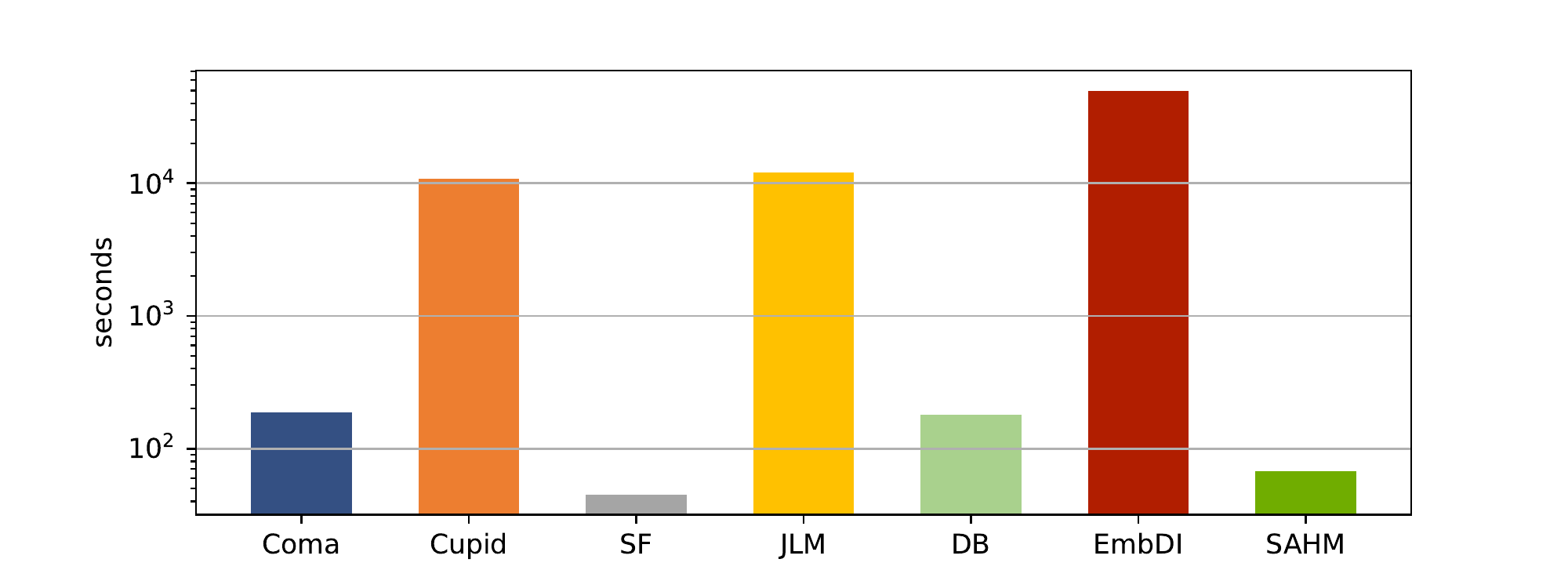}}
	
	\subfloat[\label{fig:runtime_det}]{%
		\includegraphics[width=\linewidth,height =1.7in]{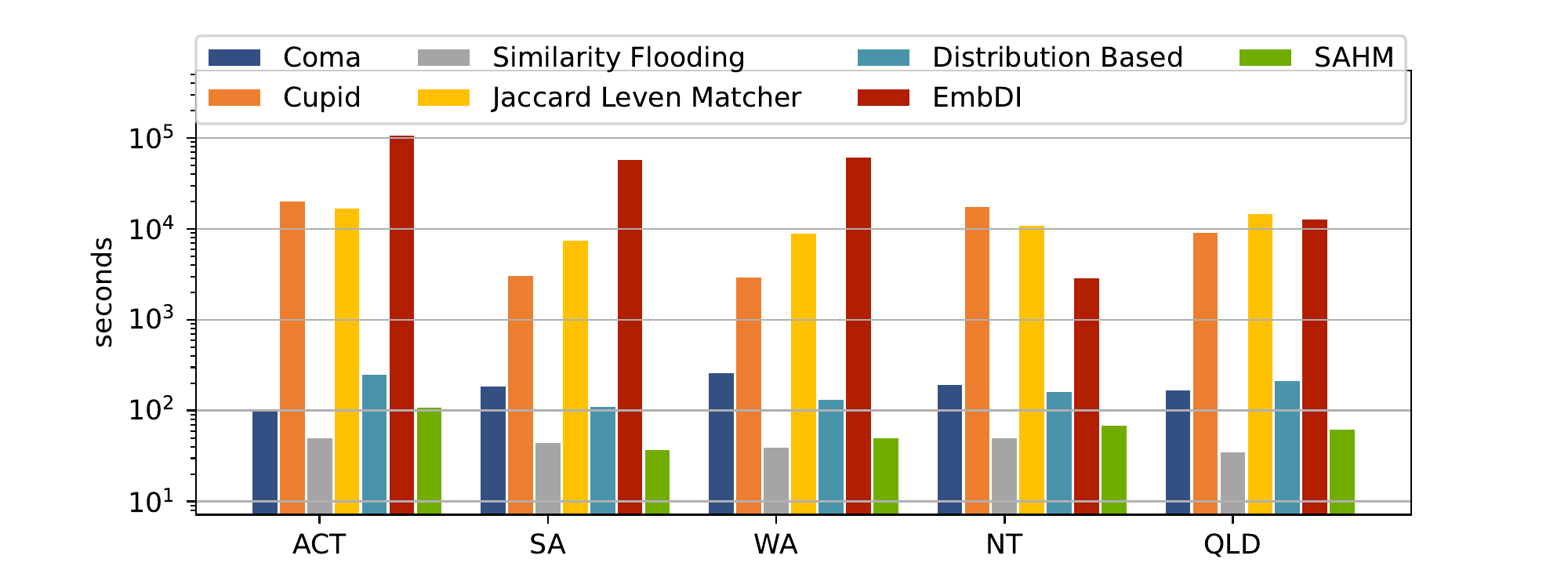}}%

	\caption{Efficiency analysis. (a) Average runtime for all state datasets for each respective matching algorithm (b) Runtime results for each state and corresponding matching algorithm.} 
	\label{fig:efficiency_results}
\end{figure}

In this experiment, we compared SAHM's efficiency  with the five other schema matching algorithms by calculating the runtimes using the proprietary NVIS datasets. Figure \ref{fig:efficiency_results} is the results of our efficiency experiments in log scale. In Figure \ref{fig:runtime_avg}, which is the average runtime across all datasets for each algorithm,  we observe that SAHM is several orders of magnitude more efficient than most of the existing state-of-the-art techniques. It is only Similirity Flooding (SF) which is marginally more efficient. Further analysis of the runtime results in \ref{fig:runtime_det}, which is the individual runtime results for each dataset, shows that for each state dataset, SAHM and SF are the most efficient algorithms. SF is more efficient compared to SAHM becase SF is purely a schema based matcher whereas SAHM uses a combination of both schema and instance level matching. We also observe that EmbDI takes the most time because EmbDI has a bottleneck in its random walk generation and as the number of instances grow, it does not scale effectively. Further, EmbDI depends on word-embeddings which adds to its computational costs. It is worth noting that while SF is the only matching algorithm that is marginally more efficient than SAHM, SF's efficiency is at a considerable cost of effectiveness. That is, the effectiveness of SF is not competitive as illustrated in Figure \ref{fig:results}.

\subsection{SAHM performance based on number of recommendations} 
\begin{figure}[t]

		\includegraphics[width=\linewidth,height =1.7in]{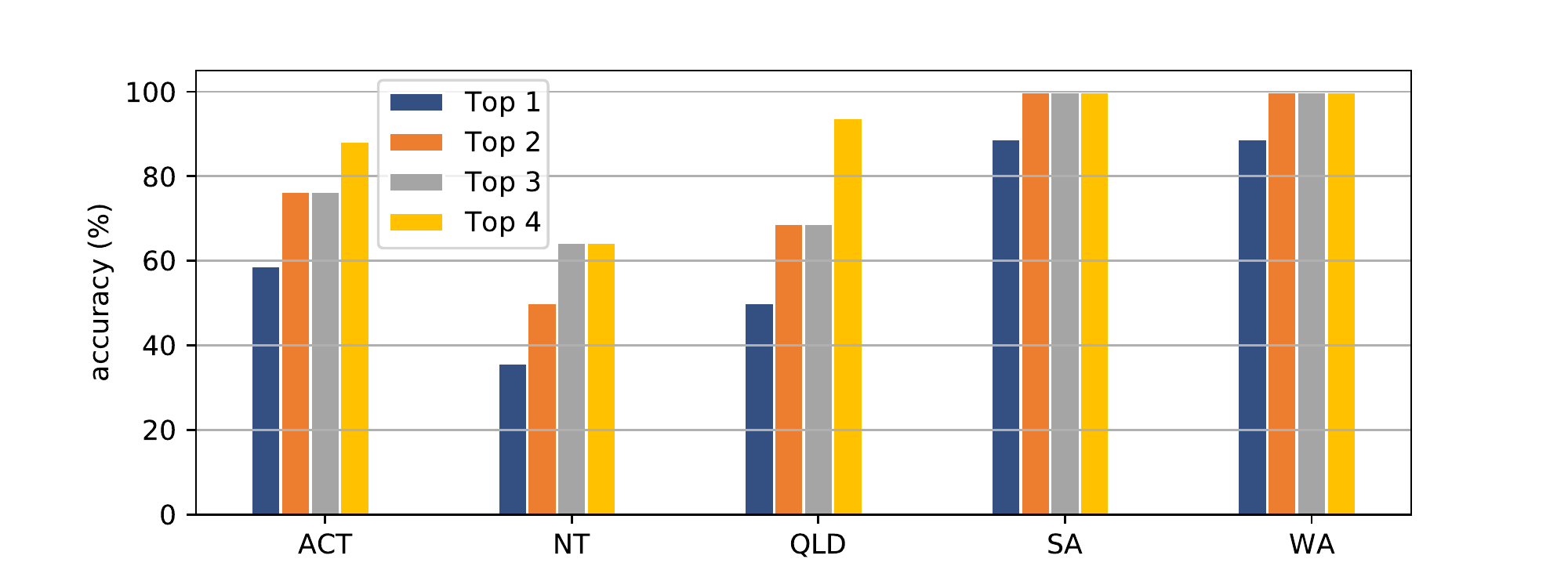}

	\caption{Percentage accuracy of state datasets based on schema matching suggestions.} 
	\label{fig:figb}
\end{figure}

In this experiment, we aim to show the influence of the parameter \emph{Top\_*} on the accuracy of SAHM.
The results are presented in Figure~\ref{fig:figb}. It is observed that, for all datasets, the performance of \emph{Top\_1} is satisfactory but increases greatly as the parameter \emph{Top\_* } is increased to \emph{Top\_4}. Our analysis in Figure~\ref{fig:figb} shows that the accuracy of the top four suggestions ranged between 88\% and 100\%  for four state datasets \emph{i.e.} ACT, QLD, SA and WA. The NT dataset however, achieves at most 64\% accuracy. As previously observed, the NT dataset performs poorest because the attribute names of the NT dataset have an unusual naming convention. 

\begin{table}[t!]
\smaller
\centering
\caption{ Impact of number of recommendations on the accuracy of individual components}
\label{tab:in_new}
\begin{tabular}{|l|l|p{1cm}|p{1cm}|p{1cm}|p{1cm}|p{1cm}|p{1cm}|}
\hline
\multicolumn{2}{|l|}{Individual Component} & & WA(\%) & SA(\%) & QLD(\%) & NT(\%) & ACT(\%)\\
\hline
\multicolumn{2}{|l|}{\multirow{4}{*}{\begin{tabular}[c]{@{}l@{}}Linguistic Similarity \\Component\end{tabular}}} & Top\_1 & 66.66 & 77.77 & 37.5 & 14.29 & 35.29 \\ \cline{3-8} 
\multicolumn{2}{|l|}{} & Top\_2 & 66.66 & 88.88 & 50 & 57.15 & 70.58 \\ \cline{3-8} 
\multicolumn{2}{|l|}{} & Top\_3 & 66.66 & 99.99 & 62.5 & 57.15 & 76.46 \\ \cline{3-8} 
\multicolumn{2}{|l|}{} & Top\_4 & 66.66 & 99.99 & 75 & 57.15 & 82.34 \\ \hline
\multicolumn{2}{|l|}{\multirow{4}{*}{Univariate Component}} & Top\_1 & 55.55 & 44.44 & 25 & 21.43 & 35.29 \\ \cline{3-8} 
\multicolumn{2}{|l|}{} & Top\_2 & 66.66 & 44.44 & 43.75 & 21.43 & 64.7 \\ \cline{3-8} 
\multicolumn{2}{|l|}{} & Top\_3 & 66.66 & 44.44 & 56.25 & 21.43 & 76.46 \\ \cline{3-8} 
\multicolumn{2}{|l|}{} & Top\_4 & 77.77 & 55.55 & 56.25 & 28.57 & 76.46 \\ \hline
\multirow{8}{*}{\rotatebox[origin=c]{90}{Multivariate Component}} 
 & \multirow{4}{*}{\begin{tabular}[c]{@{}l@{}}Randomly Matched \\Pair\end{tabular}} & Top\_1 & 55.55 & 22.22 & 18.75 & 7.14 & 0 \\ \cline{3-8} 
 &  & Top\_2 & 88.88 & 22.22 & 25 & 14.28 & 0 \\ \cline{3-8} 
 &  & Top\_3 & 88.88 & 33.33 & 37.5 & 28.57 & 5.88 \\ \cline{3-8} 
 &  & Top\_4 & 88.88 & 33.33 & 43.75 & 35.71 & 17.64 \\ \cline{2-8} 
 & \multirow{4}{*}{\begin{tabular}[c]{@{}l@{}}Known Matched \\Pair\end{tabular}} & Top\_1 & 66.66 & 22.22 & 18.75 & 21.43 & 41.17 \\ \cline{3-8} 
 &  & Top\_2 & 88.88 & 44.44 & 31.25 & 42.86 & 47.05 \\ \cline{3-8} 
 &  & Top\_3 & 88.88 & 55.55 & 37.5 & 57.15 & 52.93 \\ \cline{3-8} 
 &  & Top\_4 & 88.88 & 66.66 & 50 & 64.29 & 58.81 \\ \hline
\end{tabular}%
\end{table}
\normalfont
\subsection{Performance of individual components of SAHM }
In this experiment we aim to demonstrate the effectiveness of the individual components in SAHM and the impact of the number of recommendations on the individual matchers.
This is shown in Table \ref{tab:in_new}. We remind the reader that in the multivariate statistical matching, a known true matched pair of the source and destination attribute is used as the basis for calculating the other matches. In the table, Multivariate (randomly matched pair), represents the scenario where the user does not have any true matching pairs of the destination and source attributes, thus a random pair is chosen. Multivariate (known matched pair), represents the scenario where the user has one true matching pair of the destination and source attributes. 

The table shows the results for Linguistic Similarity Matching, Univariate Statistical Matching, and Multivariate Statistical Matching. Note that Domain Knowledge Matching was not used since the results would be trivial. 
From the table, we notice that in all cases, as the number of recommendation increases (\emph{i.e. Top\_*}), the accuracy improves monotonically.
It is interesting to note that, the \emph{Top\_1} of the Linguistic Similarity for WA dataset is 66.66\% and this does not change as the number of recommendations are increased. This value reflects that if we only consider Linguistic Similarity component in the framework for WA dataset, then schema matching accuracy for the dataset will always be 66.66\%. Clearly, for WA dataset, the multivariate component has the best performance. Further, we observe that for the SA dataset, the Linguistic Similarity is the most significant component. Strong performances of Linguistic Similarity can also be seen for QLD and ACT datasets. In NT dataset however, the Multivariate Matcher (Known matched Pair) yields much better results than Linguistic Similarity Matching. This shows that SAHM exploits the suitability of different matchers for different datasets and thus, benefits from the ensemble approach.

\section{Conclusion}
In this work, a semi-automated database schema matching framework called SAHM for matching data from heterogeneous data sources has been developed. SAHM is a two tiered framework comprising of schema level matching and instance level matching. Each tier is an ensemble of techniques which together enhances the effectiveness of the matching framework. Further, a novel multivariate statistical matching technique has been introduced, which is demonstrated to enhance existing schema matching techniques. SAHM has been evaluated on proprietary NVIS datasets and its performance has been shown to be much better than existing state-of-the-art techniques. The results show that, by leveraging domain knowledge in our novel SAHM framework, the effectiveness of schema matching can be increased tremendously. At the same time, in scenarios where domain knowledge may be lacking, SAHM still yields reasonably better results. In the future, we intend to study how the proposed schema matching framework can enhance utility in downstream machine learning applications. Furthermore, other similarity metrics such as semantic similarity and embedding based similarity approaches will be investigated. 

\section*{Acknowledgement}
This research was funded by the Department of Climate Change, Energy, the Environment and Water (DCCEEW), Australia formerly Department of Agriculture, Water and Environment (DAWE), and Data61 CSIRO, Australia.


\begin{thebibliography}{32}
\expandafter\ifx\csname natexlab\endcsname\relax\def\natexlab#1{#1}\fi
\providecommand{\url}[1]{\texttt{#1}}
\providecommand{\href}[2]{#2}
\providecommand{\path}[1]{#1}
\providecommand{\DOIprefix}{doi:}
\providecommand{\ArXivprefix}{arXiv:}
\providecommand{\URLprefix}{URL: }
\providecommand{\Pubmedprefix}{pmid:}
\providecommand{\doi}[1]{\href{http://dx.doi.org/#1}{\path{#1}}}
\providecommand{\Pubmed}[1]{\href{pmid:#1}{\path{#1}}}
\providecommand{\bibinfo}[2]{#2}
\ifx\xfnm\relax \def\xfnm[#1]{\unskip,\space#1}\fi
\bibitem[{Aumueller et~al.(2005)Aumueller, Do, Massmann \& Rahm}]{comapp1}
\bibinfo{author}{Aumueller, D.}, \bibinfo{author}{Do, H.-H.},
  \bibinfo{author}{Massmann, S.}, \& \bibinfo{author}{Rahm, E.}
  (\bibinfo{year}{2005}).
\newblock \bibinfo{title}{Schema and ontology matching with coma++}.
\newblock In {\it \bibinfo{booktitle}{Proceedings of the 2005 ACM SIGMOD
  international conference on Management of data}\/} (pp.
  \bibinfo{pages}{906--908}).
\bibitem[{Bilke \& Naumann(2005)}]{dumas}
\bibinfo{author}{Bilke, A.}, \& \bibinfo{author}{Naumann, F.}
  (\bibinfo{year}{2005}).
\newblock \bibinfo{title}{Schema matching using duplicates}.
\newblock In {\it \bibinfo{booktitle}{21st International Conference on Data
  Engineering (ICDE'05)}\/} (pp. \bibinfo{pages}{69--80}).
\newblock \bibinfo{organization}{IEEE}.
\bibitem[{Cappuzzo et~al.(2021)Cappuzzo, Papotti \& Thirumuruganathan}]{embdi}
\bibinfo{author}{Cappuzzo, R.}, \bibinfo{author}{Papotti, P.}, \&
  \bibinfo{author}{Thirumuruganathan, S.} (\bibinfo{year}{2021}).
\newblock \bibinfo{title}{Embdi: Generating embeddings for relational data
  integration}.
\newblock In {\it \bibinfo{booktitle}{SEBD 2021: The 29th Italian Symposium on
  Advanced Database Systems}\/}.
\bibitem[{Chaudhuri et~al.(2006)Chaudhuri, Ganti \& Kaushik}]{statisticalsm}
\bibinfo{author}{Chaudhuri, S.}, \bibinfo{author}{Ganti, V.}, \&
  \bibinfo{author}{Kaushik, R.} (\bibinfo{year}{2006}).
\newblock \bibinfo{title}{A primitive operator for similarity joins in data
  cleaning}.
\newblock In {\it \bibinfo{booktitle}{22nd International Conference on Data
  Engineering (ICDE'06)}\/} (pp. \bibinfo{pages}{5--5}).
\newblock \bibinfo{organization}{IEEE}.
\bibitem[{Chen et~al.(2018)Chen, Golshan, Halevy, Tan \& Doan}]{fm}
\bibinfo{author}{Chen, C.}, \bibinfo{author}{Golshan, B.},
  \bibinfo{author}{Halevy, A.~Y.}, \bibinfo{author}{Tan, W.-C.}, \&
  \bibinfo{author}{Doan, A.} (\bibinfo{year}{2018}).
\newblock \bibinfo{title}{Biggorilla: An open-source ecosystem for data
  preparation and integration.}
\newblock {\it \bibinfo{journal}{IEEE Data Eng. Bull.}\/},  {\it
  \bibinfo{volume}{41}\/}, \bibinfo{pages}{10--22}.
\bibitem[{Coen \& Xue(2009)}]{smwithdd}
\bibinfo{author}{Coen, G.}, \& \bibinfo{author}{Xue, P.}
  (\bibinfo{year}{2009}).
\newblock \bibinfo{title}{Schema-matching with data dictionaries}.
\newblock In {\it \bibinfo{booktitle}{International Conference on Application
  of Natural Language to Information Systems}\/} (pp. \bibinfo{pages}{62--78}).
\newblock \bibinfo{organization}{Springer}.
\bibitem[{Cruz et~al.(2009)Cruz, Antonelli \& Stroe}]{am}
\bibinfo{author}{Cruz, I.~F.}, \bibinfo{author}{Antonelli, F.~P.}, \&
  \bibinfo{author}{Stroe, C.} (\bibinfo{year}{2009}).
\newblock \bibinfo{title}{Agreementmaker: efficient matching for large
  real-world schemas and ontologies}.
\newblock {\it \bibinfo{journal}{Proceedings of the VLDB Endowment}\/},  {\it
  \bibinfo{volume}{2}\/}, \bibinfo{pages}{1586--1589}.
\bibitem[{Do \& Rahm(2002)}]{coma}
\bibinfo{author}{Do, H.-H.}, \& \bibinfo{author}{Rahm, E.}
  (\bibinfo{year}{2002}).
\newblock \bibinfo{title}{Coma—a system for flexible combination of schema
  matching approaches}.
\newblock In {\it \bibinfo{booktitle}{VLDB'02: Proceedings of the 28th
  International Conference on Very Large Databases}\/} (pp.
  \bibinfo{pages}{610--621}).
\newblock \bibinfo{organization}{Elsevier}.
\bibitem[{Do \& Rahm(2007)}]{comapp2}
\bibinfo{author}{Do, H.-H.}, \& \bibinfo{author}{Rahm, E.}
  (\bibinfo{year}{2007}).
\newblock \bibinfo{title}{Matching large schemas: Approaches and evaluation}.
\newblock {\it \bibinfo{journal}{Information Systems}\/},  {\it
  \bibinfo{volume}{32}\/}, \bibinfo{pages}{857--885}.
\bibitem[{Dongming \& Guohua(2010)}]{i1}
\bibinfo{author}{Dongming, Z.}, \& \bibinfo{author}{Guohua, L.}
  (\bibinfo{year}{2010}).
\newblock \bibinfo{title}{Esm: An extended schema matching method between
  database schemas}.
\newblock In {\it \bibinfo{booktitle}{2010 International Conference on
  E-Business and E-Government}\/} (pp. \bibinfo{pages}{3996--4000}).
\newblock \bibinfo{organization}{IEEE}.
\bibitem[{Fernandez et~al.(2018)Fernandez, Mansour, Qahtan, Elmagarmid, Ilyas,
  Madden, Ouzzani, Stonebraker \& Tang}]{semprop}
\bibinfo{author}{Fernandez, R.~C.}, \bibinfo{author}{Mansour, E.},
  \bibinfo{author}{Qahtan, A.~A.}, \bibinfo{author}{Elmagarmid, A.},
  \bibinfo{author}{Ilyas, I.}, \bibinfo{author}{Madden, S.},
  \bibinfo{author}{Ouzzani, M.}, \bibinfo{author}{Stonebraker, M.}, \&
  \bibinfo{author}{Tang, N.} (\bibinfo{year}{2018}).
\newblock \bibinfo{title}{Seeping semantics: Linking datasets using word
  embeddings for data discovery}.
\newblock In {\it \bibinfo{booktitle}{2018 IEEE 34th International Conference
  on Data Engineering (ICDE)}\/} (pp. \bibinfo{pages}{989--1000}).
\newblock \bibinfo{organization}{IEEE}.
\bibitem[{Gal et~al.(2019)Gal, Roitman \& Shraga}]{learning_to_rerank}
\bibinfo{author}{Gal, A.}, \bibinfo{author}{Roitman, H.}, \&
  \bibinfo{author}{Shraga, R.} (\bibinfo{year}{2019}).
\newblock \bibinfo{title}{Learning to rerank schema matches}.
\newblock {\it \bibinfo{journal}{IEEE Transactions on Knowledge and Data
  Engineering}\/}, .
\bibitem[{Haslem et~al.(2010)Haslem, Callister, Avitabile, Griffioen, Kelly,
  Nimmo, Spence-Bailey, Taylor, Watson, Brown et~al.}]{impnvis2}
\bibinfo{author}{Haslem, A.}, \bibinfo{author}{Callister, K.~E.},
  \bibinfo{author}{Avitabile, S.~C.}, \bibinfo{author}{Griffioen, P.~A.},
  \bibinfo{author}{Kelly, L.~T.}, \bibinfo{author}{Nimmo, D.~G.},
  \bibinfo{author}{Spence-Bailey, L.~M.}, \bibinfo{author}{Taylor, R.~S.},
  \bibinfo{author}{Watson, S.~J.}, \bibinfo{author}{Brown, L.} et~al.
  (\bibinfo{year}{2010}).
\newblock \bibinfo{title}{A framework for mapping vegetation over broad spatial
  extents: a technique to aid land management across jurisdictional
  boundaries}.
\newblock {\it \bibinfo{journal}{Landscape and Urban Planning}\/},  {\it
  \bibinfo{volume}{97}\/}, \bibinfo{pages}{296--305}.
\bibitem[{Hu et~al.(2008)Hu, Qu \& Cheng}]{fa}
\bibinfo{author}{Hu, W.}, \bibinfo{author}{Qu, Y.}, \& \bibinfo{author}{Cheng,
  G.} (\bibinfo{year}{2008}).
\newblock \bibinfo{title}{Matching large ontologies: A divide-and-conquer
  approach}.
\newblock {\it \bibinfo{journal}{Data \& Knowledge Engineering}\/},  {\it
  \bibinfo{volume}{67}\/}, \bibinfo{pages}{140--160}.
\bibitem[{Kang \& Naughton(2003)}]{stringmatcher2}
\bibinfo{author}{Kang, J.}, \& \bibinfo{author}{Naughton, J.~F.}
  (\bibinfo{year}{2003}).
\newblock \bibinfo{title}{On schema matching with opaque column names and data
  values}.
\newblock In {\it \bibinfo{booktitle}{Proceedings of the 2003 ACM SIGMOD
  international conference on Management of data}\/} (pp.
  \bibinfo{pages}{205--216}).
\bibitem[{Klein et~al.(2009)Klein, Wilson, Watts, Stein, Carwardine, Mackey \&
  Possingham}]{impnvis3}
\bibinfo{author}{Klein, C.~J.}, \bibinfo{author}{Wilson, K.~A.},
  \bibinfo{author}{Watts, M.}, \bibinfo{author}{Stein, J.},
  \bibinfo{author}{Carwardine, J.}, \bibinfo{author}{Mackey, B.}, \&
  \bibinfo{author}{Possingham, H.~P.} (\bibinfo{year}{2009}).
\newblock \bibinfo{title}{Spatial conservation prioritization inclusive of
  wilderness quality: A case study of australia’s biodiversity}.
\newblock {\it \bibinfo{journal}{Biological Conservation}\/},  {\it
  \bibinfo{volume}{142}\/}, \bibinfo{pages}{1282--1290}.
\bibitem[{Koutras et~al.(2021)Koutras, Siachamis, Ionescu, Psarakis, Brons,
  Fragkoulis, Lofi, Bonifati \& Katsifodimos}]{valentine}
\bibinfo{author}{Koutras, C.}, \bibinfo{author}{Siachamis, G.},
  \bibinfo{author}{Ionescu, A.}, \bibinfo{author}{Psarakis, K.},
  \bibinfo{author}{Brons, J.}, \bibinfo{author}{Fragkoulis, M.},
  \bibinfo{author}{Lofi, C.}, \bibinfo{author}{Bonifati, A.}, \&
  \bibinfo{author}{Katsifodimos, A.} (\bibinfo{year}{2021}).
\newblock \bibinfo{title}{Valentine: Evaluating matching techniques for dataset
  discovery}.
\newblock In {\it \bibinfo{booktitle}{2021 IEEE 37th International Conference
  on Data Engineering (ICDE)}\/} (pp. \bibinfo{pages}{468--479}).
\newblock \bibinfo{organization}{IEEE}.
\bibitem[{Levenshtein et~al.(1966)}]{levenshtein}
\bibinfo{author}{Levenshtein, V.~I.} et~al. (\bibinfo{year}{1966}).
\newblock \bibinfo{title}{Binary codes capable of correcting deletions,
  insertions, and reversals}.
\newblock In {\it \bibinfo{booktitle}{Soviet physics doklady}\/} (pp.
  \bibinfo{pages}{707--710}).
\newblock \bibinfo{organization}{Soviet Union} volume~\bibinfo{volume}{10}.
\bibitem[{Madhavan et~al.(2001)Madhavan, Bernstein \& Rahm}]{cupid}
\bibinfo{author}{Madhavan, J.}, \bibinfo{author}{Bernstein, P.~A.}, \&
  \bibinfo{author}{Rahm, E.} (\bibinfo{year}{2001}).
\newblock \bibinfo{title}{Generic schema matching with cupid}.
\newblock In {\it \bibinfo{booktitle}{vldb}\/} (pp. \bibinfo{pages}{49--58}).
\newblock \bibinfo{organization}{Citeseer} volume~\bibinfo{volume}{1}.
\bibitem[{Melnik et~al.(2002)Melnik, Garcia-Molina \& Rahm}]{comapp3}
\bibinfo{author}{Melnik, S.}, \bibinfo{author}{Garcia-Molina, H.}, \&
  \bibinfo{author}{Rahm, E.} (\bibinfo{year}{2002}).
\newblock \bibinfo{title}{Similarity flooding: A versatile graph matching
  algorithm and its application to schema matching}.
\newblock In {\it \bibinfo{booktitle}{Proceedings 18th International Conference
  on Data Engineering}\/} (pp. \bibinfo{pages}{117--128}).
\newblock \bibinfo{organization}{IEEE}.
\bibitem[{Monge et~al.(1996)Monge, Elkan et~al.}]{monge}
\bibinfo{author}{Monge, A.~E.}, \bibinfo{author}{Elkan, C.} et~al.
  (\bibinfo{year}{1996}).
\newblock \bibinfo{title}{The field matching problem: Algorithms and
  applications.}
\newblock In {\it \bibinfo{booktitle}{Kdd}\/} (pp. \bibinfo{pages}{267--270}).
\newblock volume~\bibinfo{volume}{2}.
\bibitem[{{NVIS Technical Working Group}(2017)}]{nvismanual}
\bibinfo{author}{{NVIS Technical Working Group}} (\bibinfo{year}{2017}).
\newblock \bibinfo{title}{Australian vegetation attribute manual: National
  vegetation information system, version 7.0}.
\newblock {\it \bibinfo{journal}{Department of the Environment and Energy,
  Canberra}\/}, .
\bibitem[{Rahm \& Bernstein(2001)}]{stringmatcher1}
\bibinfo{author}{Rahm, E.}, \& \bibinfo{author}{Bernstein, P.~A.}
  (\bibinfo{year}{2001}).
\newblock \bibinfo{title}{A survey of approaches to automatic schema matching}.
\newblock {\it \bibinfo{journal}{the VLDB Journal}\/},  {\it
  \bibinfo{volume}{10}\/}, \bibinfo{pages}{334--350}.
\bibitem[{Rahman \& Islam(2016)}]{discretization}
\bibinfo{author}{Rahman, M.~G.}, \& \bibinfo{author}{Islam, M.~Z.}
  (\bibinfo{year}{2016}).
\newblock \bibinfo{title}{Discretization of continuous attributes through low
  frequency numerical values and attribute interdependency}.
\newblock {\it \bibinfo{journal}{Expert Systems with Applications}\/},  {\it
  \bibinfo{volume}{45}\/}, \bibinfo{pages}{410--423}.
\bibitem[{Salton \& McGill(1983)}]{tfidf}
\bibinfo{author}{Salton, G.}, \& \bibinfo{author}{McGill, M.~J.}
  (\bibinfo{year}{1983}).
\newblock {\it \bibinfo{title}{Introduction to modern information
  retrieval}\/}.
\newblock \bibinfo{publisher}{mcgraw-hill}.
\bibitem[{Seligman et~al.(2010)Seligman, Mork, Halevy, Smith, Carey, Chen,
  Wolf, Madhavan, Kannan \& Burdick}]{hm}
\bibinfo{author}{Seligman, L.}, \bibinfo{author}{Mork, P.},
  \bibinfo{author}{Halevy, A.}, \bibinfo{author}{Smith, K.},
  \bibinfo{author}{Carey, M.~J.}, \bibinfo{author}{Chen, K.},
  \bibinfo{author}{Wolf, C.}, \bibinfo{author}{Madhavan, J.},
  \bibinfo{author}{Kannan, A.}, \& \bibinfo{author}{Burdick, D.}
  (\bibinfo{year}{2010}).
\newblock \bibinfo{title}{Openii: an open source information integration
  toolkit}.
\newblock In {\it \bibinfo{booktitle}{Proceedings of the 2010 ACM SIGMOD
  International Conference on Management of data}\/} (pp.
  \bibinfo{pages}{1057--1060}).
\bibitem[{Shrestha et~al.(2019)Shrestha, Tran, Bhattarai, Pusey \& Aygun}]{pc}
\bibinfo{author}{Shrestha, M.}, \bibinfo{author}{Tran, T.~X.},
  \bibinfo{author}{Bhattarai, B.}, \bibinfo{author}{Pusey, M.~L.}, \&
  \bibinfo{author}{Aygun, R.~S.} (\bibinfo{year}{2019}).
\newblock \bibinfo{title}{Schema matching and data integration with consistent
  naming on protein crystallization screens}.
\newblock {\it \bibinfo{journal}{IEEE/ACM transactions on computational biology
  and bioinformatics}\/},  {\it \bibinfo{volume}{17}\/},
  \bibinfo{pages}{2074--2085}.
\bibitem[{Smith et~al.(2009)Smith, Morse, Mork, Li, Rosenthal, Allen, Seligman
  \& Wolf}]{impofsm}
\bibinfo{author}{Smith, K.}, \bibinfo{author}{Morse, M.},
  \bibinfo{author}{Mork, P.}, \bibinfo{author}{Li, M.},
  \bibinfo{author}{Rosenthal, A.}, \bibinfo{author}{Allen, D.},
  \bibinfo{author}{Seligman, L.}, \& \bibinfo{author}{Wolf, C.}
  (\bibinfo{year}{2009}).
\newblock \bibinfo{title}{The role of schema matching in large enterprises}.
\newblock {\it \bibinfo{journal}{arXiv preprint arXiv:0909.1771}\/}, .
\bibitem[{Thackway et~al.(2007)Thackway, Lee, Donohue, Keenan \&
  Wood}]{impnvis1}
\bibinfo{author}{Thackway, R.}, \bibinfo{author}{Lee, A.},
  \bibinfo{author}{Donohue, R.}, \bibinfo{author}{Keenan, R.~J.}, \&
  \bibinfo{author}{Wood, M.} (\bibinfo{year}{2007}).
\newblock \bibinfo{title}{Vegetation information for improved natural resource
  management in australia}.
\newblock {\it \bibinfo{journal}{Landscape and Urban Planning}\/},  {\it
  \bibinfo{volume}{79}\/}, \bibinfo{pages}{127--136}.
\bibitem[{Zhang et~al.(2010)Zhang, Hadjieleftheriou, Ooi, Procopiuc \&
  Srivastava}]{emd}
\bibinfo{author}{Zhang, M.}, \bibinfo{author}{Hadjieleftheriou, M.},
  \bibinfo{author}{Ooi, B.~C.}, \bibinfo{author}{Procopiuc, C.~M.}, \&
  \bibinfo{author}{Srivastava, D.} (\bibinfo{year}{2010}).
\newblock \bibinfo{title}{On multi-column foreign key discovery}.
\newblock {\it \bibinfo{journal}{Proceedings of the VLDB Endowment}\/},  {\it
  \bibinfo{volume}{3}\/}, \bibinfo{pages}{805--814}.
\bibitem[{Zhang et~al.(2011{\natexlab{a}})Zhang, Hadjieleftheriou, Ooi,
  Procopiuc \& Srivastava}]{automatic_discovery_of_attribute}
\bibinfo{author}{Zhang, M.}, \bibinfo{author}{Hadjieleftheriou, M.},
  \bibinfo{author}{Ooi, B.~C.}, \bibinfo{author}{Procopiuc, C.~M.}, \&
  \bibinfo{author}{Srivastava, D.} (\bibinfo{year}{2011}{\natexlab{a}}).
\newblock \bibinfo{title}{Automatic discovery of attributes in relational
  databases}.
\newblock In {\it \bibinfo{booktitle}{Proceedings of the 2011 ACM SIGMOD
  International Conference on Management of data}\/} (pp.
  \bibinfo{pages}{109--120}).
\bibitem[{Zhang et~al.(2011{\natexlab{b}})Zhang, Hadjieleftheriou, Ooi,
  Procopiuc \& Srivastava}]{distributionbased}
\bibinfo{author}{Zhang, M.}, \bibinfo{author}{Hadjieleftheriou, M.},
  \bibinfo{author}{Ooi, B.~C.}, \bibinfo{author}{Procopiuc, C.~M.}, \&
  \bibinfo{author}{Srivastava, D.} (\bibinfo{year}{2011}{\natexlab{b}}).
\newblock \bibinfo{title}{Automatic discovery of attributes in relational
  databases}.
\newblock In {\it \bibinfo{booktitle}{Proceedings of the 2011 ACM SIGMOD
  International Conference on Management of data}\/} (pp.
  \bibinfo{pages}{109--120}).

\end{thebibliography}

\end{document}